\documentclass[conference]{IEEEtran}
\usepackage{cite}
\usepackage{amsmath,amssymb,amsfonts}
\usepackage{algorithmic}
\usepackage{graphicx}
\usepackage{textcomp}
\usepackage[numbers]{natbib}  

\DeclareUnicodeCharacter{221A}{ }

\usepackage{subfig}
\usepackage[table,xcdraw]{xcolor} 
\usepackage{floatrow}
\usepackage{multirow}

\usepackage{xcolor}
\usepackage{url}

\usepackage{soul} 

\usepackage{refcount} %

\usepackage{lscape}

\def\BibTeX{{\rm B\kern-.05em{\sc i\kern-.025em b}\kern-.08em
    T\kern-.1667em\lower.7ex\hbox{E}\kern-.125emX}}

\newtheorem{assump}{Assumption}
\newtheorem{hyp}{Hypothesis}
    
\begin{document}

\title{To incentivize or not: Impact of blockchain-based cryptoeconomic tokens on human information sharing behavior}

\author{\IEEEauthorblockN{Mark C. Ballandies}
	\IEEEauthorblockA{Computational Social Science \\
		ETH Zurich\\
		Zurich, Switzerland\\ 
		bcmark@protonmail.com}
}

\maketitle

\begin{abstract}
Cryptoeconomic incentives in the form of blockchain-based tokens are seen as an enabler of the sharing economy that could shift society towards greater sustainability. 
Nevertheless, knowledge of the impact of these tokens on human sharing behavior is still limited and this poses a
challenge to the design of effective cryptoeconomic incentives.
This study applies the theory of self-determination to investigate the impact of such tokens on human behavior in an information-sharing scenario. 
By utilizing an experimental methodology in the form of a randomized control trial with a 2x2 factorial design involving 132 participants, the effects of two token incentives on human information-sharing behavior are analyzed.
Individuals obtain these tokens in exchange for their shared information. Based on the collected tokens, individuals receive a monetary payment and build reputation. 
Besides investigating the effect of these incentives on the quantity of shared information, the study includes quality characteristics of the information, such as \textit{accuracy} and \textit{contextualization}. The focus on quantity while excluding quality has been identified as a limitation in previous work.
In addition to confirming previously known effects such as a crowding-out of intrinsic motivation by incentives, which also exists for blockchain-based tokens, the findings of this paper point to a hitherto unreported \textit{interaction effect} between multiple tokens when applied simultaneously.
The findings are critically discussed and put into the context of recent work and ethical considerations. 
The theory-based-empirical study is of interest to those investigating the effect of cryptoeconomic tokens or digital currencies on human behavior and supports the community in the design of effective personalized incentives for sharing economies. 
\end{abstract}

%

\section{Introduction}
Cryptoeconomic incentives in the form of blockchain-based tokens are seen as an enabler of the sharing economy~\cite{pazaitis2017blockchain,ferraro2018distributed} that could shift society toward greater sustainability~\cite{heinrichs2013sharing,fanitabasi2021self}. One of the resources that is shared in these economies is information~\cite{richter2019data,raweewan2018information,nonaka1994dynamic}, which due to its growing utilization in data-intensive technologies~\cite{bennati2018machine} is becoming increasingly important~\cite{helbing2022socio, economist2017world}. This has resulted in the collection of large data sets by organizations~\cite{cai2015challenges}. Nevertheless, in this age of vast data quantities, obtaining high-quality information is a challenge~\cite{cai2015challenges,gao2016big} (e.g. the accuracy of the collected data is low). Moreover, because organizations collect massive amounts of unstructured data, such as customer behavior (product choices and sleeping patterns), opinions (e.g. Facebook likes), medical health records, or IoT data, the amount of data collected exceeds the processing power available to analyze it~\cite{helbing2015digital}, possibly resulting in sampling biases. Furthermore, these "Big Data" approaches often involve the danger of collapsing the complexity of entire human personalities into assumptions constructed from simple data (e.g. website clicks) and usually miss the unique domain-specific knowledge users have~\cite{lukyanenko2011citizen}. Thus, it has been suggested that information providers should structure their input in a contextualized way when sharing their data, utilizing semantic web technologies~\cite{ballandies2022improving}, such as linked data and ontologies~\cite{berners2001semantic,w32004semantic}, and evaluate the quality of information shared by other providers~\cite{paulheim2017knowledge}. 
Nevertheless, as this would require additional effort on the part of the information providers~\cite{paulheim2017knowledge}, incentives such as gamification~\cite{re2018framework}, reputation~\cite{zhou2020smart}, money~\cite{luo2019improving}, or auctions~\cite{chen2019toward} are suggested to motivate the data providers and thus improve the characteristics of the collected information. However, previous work on the incentivization of information sharing focuses on the quantity of collected information while excluding quality characteristics such as accuracy or contextualization~\cite{restuccia2016incentive}.

Increasingly, cyrptoeconomic incentives in the form of blockchain-based tokens are proposed to be awarded to participants of information-sharing communities~\cite{shrestha2018blockchain,zou2019reportcoin, makhdoom2020privysharing,jung2021mechanism,naz2019secure}. In these studies, the performance of the applied incentives is investigated using simulations~\cite{lakhani2021token}, game-theoretical methodologies~\cite{imanimehr2019token}, and case studies~\cite{hunhevicz2020incentivizing}. Nonetheless, behavioral data of users in comparison with treatment groups with and without incentivization have not been collected, which limits these approaches as the utilized models cannot be calibrated with real-world data~\cite{wittekcrypto}. Controlled experiments are therefore required that investigate the impact of these cryptoeconomic incentives on humans information-sharing behavior. In particular, such an empirical approach could assess and validate the accuracy of the utilized theoretical models~\cite{pazaitis2017blockchain}. 
Likewise, although the application of \textit{multiple} token incentives has been proposed to improve the maintenance and sharing of a common resource~\cite{hunhevicz2020incentivizing,kleineberg2021social} and has been investigated in games~\cite{imanimehr2019token} and simulations~\cite{pardi2021chemical}, the impact of simultaneously applying these incentives has not been investigated empirically in a controlled experiment. 

This paper addresses these identified gaps with the following research question:

"What is the effect of multiple blockchain-based tokens on human information-sharing behavior measured in the quantity, accuracy, and contextualization of the shared information?"

 By testing hypotheses that are informed from self-determination theory~\cite{deci1991motivational,ryan2000self} with an experimental methodology in the form of a randomized control trial utilizing a 2x2 factorial design, the impact of two types of cryptoeconomic incentives in the form of blockchain-based tokens on the information-sharing behavior of humans is investigated.


The contributions of this paper can be summarized as follows:

\begin{itemize}
    \item A conceptual impact model (Figure \ref{fig:hypotheses_model}) links cryptoeconomic incentives to human motivation and information-sharing behavior in consideration of self-determination theory~\cite{deci1991motivational,ryan2000self}.
    \item The living lab experimental methodology~\cite{pournaras2022how} is augmented with a 2x2 factorial design to investigate the impacts of blockchain-based cryptoeconomic incentives on human information-sharing behavior.
   \item Four effects of cryptoeconomic tokens on human behavior are identified: i) a hitherto unreported interaction effect between two types of cryptoeconomic tokens when applied simultaneously; ii) an internalization effect of cryptoeconomic tokens in the form of improved information-sharing behavior even after the incentivization period has ended;  iii) a crowding-out effect on intrinsic motivation when cryptoeconomic tokens are applied; iv) a time effect resulting in a variation of the impact of cryptoeconomic incentivization over time.  
    \item A novel high-quality dataset illustrates user information-sharing behavior under multiple token incentives that facilitates causal inferences about human behavior under cryptoeconomic incentivization. 
    \item The work demonstrates how self determination theory can be applied in the formulation of hypotheses and testing in Token Engineering and Token Economics.
    \item The implications of the findings for the design and engineering of one-dimensional and multi-dimensional token systems are discussed critically, taking into account ethical impacts.  
\end{itemize}
Since these contributions inform an improved construction of blockchain-based incentives, they are of relevance for the community, which is increasingly utilizing and investigating such incentives in various application domains~\cite{cai2019analysis,barreiro2019blockchain,gan2020token}. 

This paper is structured as follows: In Section \ref{sec:rel_work}, related work in information sharing is discussed. The research methodology is introduced in Section \ref{sec:research_methodology}, while Section \ref{sec:results} presents the evaluation. Section \ref{sec:discussion} summarizes the findings and discusses their implications. Finally, Section~\ref{sec:conclusion} draws the conclusion and provides an outlook for future work. 

\section{Related Work in information sharing}
\label{sec:rel_work}



\subsection{Self-determination theory and incentives}
\label{sec:std_theory}
Humans are intrinsically and extrinsically motivated to share information~\cite{osterloh2000motivation}. Intrinsic motivation refers to when people perform a task such as information sharing out of the pleasure they derive from the task itself, whereas extrinsic motivation stems from incentives, such as monetary payments, reputation gains, or punishments. When compared to extrinsic motivation for a specific task, intrinsic motivation leads to enhanced performance, persistence, creativity, learning capacity, and endurance in humans~\cite{ryan2000self} and may therefore be more important than extrinsic motivation for specific scenarios such as contributing computer code~\cite{dapp2009effects} or sharing information~\cite{palmisano2008motivational}. 

Introduced by \citet{deci2013intrinsic,deci1991motivational}, self-determination theory illustrates the conditions under which humans are intrinsically motivated to work on a task: Three innate psychological needs must be satisfied: "competence", "autonomy" and "relatedness". In particular, a feeling of competence does not enhance intrinsic motivation unless accompanied by a sense of autonomy~\cite{ryan2000self,harder2008rewards}. In this context, applying misaligned incentives may infringe on the perceived autonomy of humans and thereby reduce their intrinsic motivation~\cite{amabile1993motivational}; this is referred to as crowding-out effect~\cite{osterloh2000motivation}. 
However, competence-enhancing incentives may support intrinsic motivation ~\cite{amabile1993motivational,deci1991motivational} and this is referred to as internalization~\cite{deci1991motivational}.

\begin{figure*}[tbh!]
\begin{center}
\includegraphics[width=0.75\textwidth]{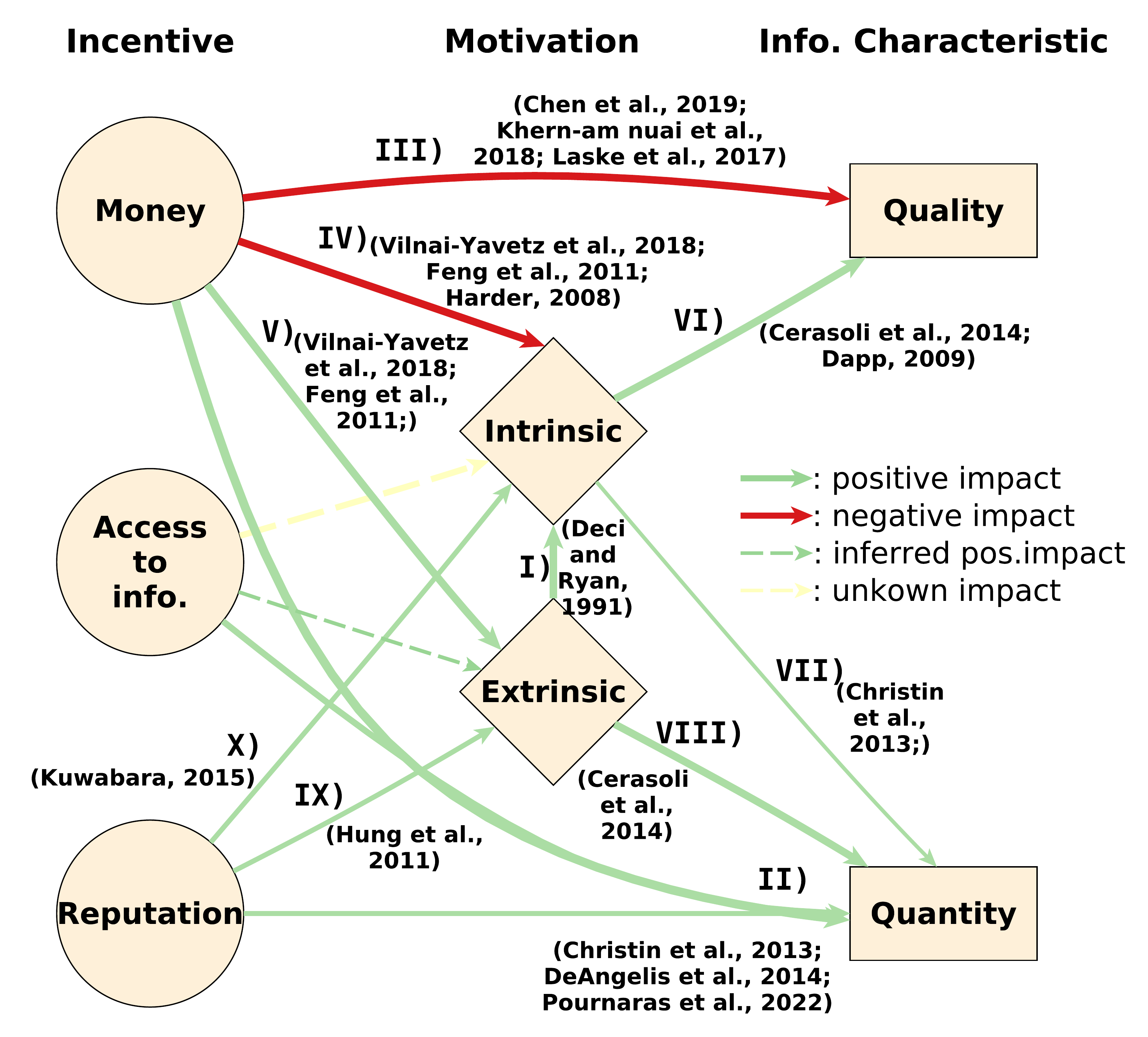}
\end{center}
\caption{Impact of incentives (money/ access/ reputation) on motivation (intrinsic/ extrinsic) and information-sharing behavior (quality/ quantity). Impacts (arrows) are derived from related work (in parentheses). }\label{fig:hypotheses_model}
\end{figure*}

Figure \ref{fig:hypotheses_model} illustrates findings from literature about the dependencies between different types of incentives, the two types of motivation, and their impact on characteristics of information. Extrinsic and intrinsic motivation are not separate systems, but influence each other~\cite{amabile1993motivational}. Extrinsic motivation can be integrated to become intrinsic motivation (Arrow I in Figure \ref{fig:hypotheses_model})~\cite{deci1991motivational}. People are extrinsically motivated to share a higher quantity of information by money~\cite{christin2013s,pournaras2022how}, the access to the collected information~\cite{christin2013s,deangelis2014systemic}, reputation~\cite{christin2013s}, and their intrinsic motivation~\cite{christin2013s} (Arrows II and VII in Figure \ref{fig:hypotheses_model}). In particular, the strongest increase in quantity is observed for money, followed by the access to information, reputation, and then intrinsic motivation~\cite{christin2013s}.
In contrast to quantity, it has been observed that the quality of shared information remains unaffected under monetary incentives when mesasuring it in terms of the prediction accuracy of stock recommendations~\cite{chen2019monetary} or is even worse than before incentivization when meassured 
by a word count index~\cite{khern2018extrinsic}, or useability (helpfulness of reviews)~\cite{khern2018extrinsic}, or quality of produced images~\cite{laske2017quantity} (Arrow III in Figure \ref{fig:hypotheses_model}).  


It has been found that monetary incentives decrease intrinsic motivation (Arrow IV in Figure \ref{fig:hypotheses_model}), while they increase extrinsic motivation (Arrow V in Figure \ref{fig:hypotheses_model})~\cite{feng2011effect,vilnai2018motivating,harder2008rewards}, which might explain the impact of monetary incentives on quality: As intrinsic motivation has been observed to predict quality (Arrow VI in Figure \ref{fig:hypotheses_model} )~\cite{cerasoli2014intrinsic,dapp2009effects} and only to a lesser extent quantity (Arrow VII in Figure \ref{fig:hypotheses_model})~\cite{christin2013s}, whereas extrinsic motivation predicts the quantity of shared information (Arrow VIII in Figure \ref{fig:hypotheses_model})~\cite{cerasoli2014intrinsic}, the use of monetary incentives would result in an increased extrinsic motivation and decreased intrinsic motivation, thereby leading to a higher quantity but lower quality of shared information. 


In contrast to monetary incentives, because they may enhance an individuals feeling of competence~\cite{amabile1993motivational,deci1991motivational}, reputation systems impact extrinsic motivation positively~\cite{hung2011influence}, while also having a positive impact on intrinsic motivation~\cite{kuwabara2015reputation} (Arrows IX and X in Figure \ref{fig:hypotheses_model}).  

Moreover, it has been found that rewarding each information-sharing action is more effective than summarized payments~\cite{yang2008knowledge}, and that small subgroups have shown moderate to strong aversion to incentives~\cite{sadler2018incentives}, which has been confirmed by Pournaras et al.~\cite{pournaras2022how}.



\begin{table*}[t] \caption{Related work that studies blockchain technology and tokens to improve information sharing. Frame: Famework; Impl.: Implemenation; Sim.: Simulation; Anly.: Analytical; Exp.: Experiment; Mult.: Multiple; Des.: Design; Imp.: Implication} \label{tab:rel_work}
\begin{tabular}{llcccccccccc} \hline \\
\textbf{ID} & \textbf{Paper}                                      & \multicolumn{2}{c}{\textbf{Artifact}} & \multicolumn{4}{c}{\textbf{Evaluation}}                          & \multicolumn{3}{c}{\textbf{Token}}              & \multicolumn{1}{l}{\textbf{Ethics}} \\
            &                                                     & \textit{Fram.}    & \textit{Impl.}   & \textit{Sim.} & \textit{Anly.} & \textit{Work.} & \textit{Exp.} & \textit{Mult.} & \textit{Des.} & \textit{Imp.} & \textit{}                           \\ \hline \\
1           & \citet{pazaitis2017blockchain}     & x                  &                  &               &                 &                &               & x               &               &               & x                                   \\
2           & \citet{shrestha2018blockchain}     & x                  &                  &               &                 &                &               &                 &               &               &                                     \\
3           & \citet{hulsemann2019walk}          & x                  &                  & x             &                 &                &               &                 &               & x             &                                     \\
4           & \citet{naz2019secure}              & x                  & x                & x             &                 &                &               &                 &               &               &                                     \\
5           & Imanimehr et. al [31]      & x                  &                  & x             & x               &                &               & x               &               & x             &                                     \\
6           & \citet{manoj2020incentive}         & x                  &                  &               &                 &                &               &                 &               &               &                                     \\
7           & \citet{hunhevicz2020incentivizing} & x                  & x                &               &                 & x              &               &                 &               &               &                                     \\
8           & \citet{zhang2020design}            & x                  & x                &               &                 &                &               &                 &               &               &                                     \\
9           & \citet{wittekcrypto}               & x                  &                  &               &                 &                &               &                 &               &               &                                     \\
10          & \citet{jaiman2021user}             & x                  & x                & x             &                 &                &               &                 &         x      &               &                                     \\
11          & \citet{jung2021mechanism}          & x                  &                  & x             & x               &                &               &                 &               &               &                                     \\
12          & \textbf{This paper}                                  & \textbf{x}         & \textbf{x}       & \textit{}     & \textit{}       & \textit{}      & \textbf{x}    & \textbf{x}      & \textbf{x}    &   \textbf{x}              & \textbf{x}              \\ \hline           
\end{tabular}
\end{table*}

\subsection{Tokens for information sharing}



Cryptoeconomic incentives in the form of blockchain-based tokens span a multi-dimensional incentive system~\cite{kleineberg2021social} enabling a differentiated pricing of a broader spectrum of externalities ~\cite{kleineberg2021social, helbing2021qualified}. This can result in the improved self-organization of society when compared to one-dimensional incentive systems such as the current monetary system~\cite{dapp2021finance,dapp2018finance}.
These tokens are defined as a \textit{"a unit of value issued within a DLT
system [or blockchain system] and which can be used as a medium of exchange or
unit of account"}~\cite{ballandies2021decrypting} and are increasingly utilized in communities to encourage the sharing of information.

Table \ref{tab:rel_work} illustrates related work that utilizes blockchain-based tokens in information-sharing scenarios.
All of these works contribute a conceptual framework about blockchain and tokens and how they can be applied to improve the information sharing in a community (Column Fram. in Table \ref{tab:rel_work}). Four of these frameworks are implemented in a software artifact (Column Impl. in Table \ref{tab:rel_work}): \citet{naz2019secure} implement a software artifact that integrates IPFS\footnote{Inter Plenatary File System, a peer-to-peer protocol for exchanging files: https://ipfs.io/ (last accessed 2022-02-07)} with a blockchain to improve the quality of shared data by incentivizing stakeholders with tokens to review the shared information. Similarly, \citet{hunhevicz2020incentivizing} use tokens to incentivize high-quality datasets in a construction process by awarding those that provide complete and accurate information. \citet{zhang2020design} use tokens in their prototype to incentivize the provision of credit data. Finally, \citet{jaiman2021user} utilize tokens as representations of ownership and access rights to data sets.
Two of these implementations are evaluated in simulations (ID 4 and 10 in Table \ref{tab:rel_work}). Moreover, three of the proposed concepts that are not implemented in a software artifact are tested in simulations (Column Sim.; ID 3, 5 and 11 in Table \ref{tab:rel_work}). \citet{hulsemann2019walk} apply an agent-based modeling approach to investigate three different types of tokens. Likewise, \citet{imanimehr2019token} investigate with a simulation the application of multiple tokens to incentivize the optimal utilization of video stream layers. Moreover, they analytically investigate their scenario with methods from game theory (Column Anly. in Table \ref{tab:rel_work}). Similarly, \citet{jung2021mechanism} utilize both methods from game theory/ mechanism design as well as simulations to evaluate their framework that improves the provision and maintenance of patient health records. \citet{hunhevicz2020incentivizing} evaluate their framework and implementation with stakeholders in a workshop (Column Work.; ID 7 in Table \ref{tab:rel_work}).
Three frameworks are not evaluated (ID 1, 2 and 6 in Table \ref{tab:rel_work}).

By utilizing tokens in their frameworks and implementations, only two of the contributions investigate the implications of introduced tokens on system properties (Column Imp.; ID 3 and 5 in Table \ref{tab:rel_work}). Moreover, only one of the contributions illustrates the token design (Column Des.; ID 10 in Table \ref{tab:rel_work}): The token is a modified ERC-721 token that has a source of value of ownership/access rights to data, its supply is uncapped and the token is transferable. Nevertheless, a standard illustration as utilized by \citet{dobler2019extension,ballandies2021finance} that would make different designs comparable has not been applied.
Moreover, the impact of a specific token design on user information-sharing behavior is not rigorously investigated with a controlled experiment (Column Exp. in Table \ref{tab:rel_work}). Assumptions thus have to be utilized in the above-mentioned simulations and analysis that limit the applicability of the findings to real-world scenarios.
Two of the works utilize multiple tokens in their application scenario (Column Mult.; ID 1 and 5 in Table \ref{tab:rel_work}): \citet{pazaitis2017blockchain} enable the setup of multiple tokens to capture the value created in different decentralised communities, and \citet{imanimehr2019token} use multiple tokens for optimal sharing of bandwidth.
Finally, despite touching on sensitive applications domains such as health or credit data, only one of the works discusses the ethical implications of their contributions and findings (Column Ethics; ID 1 in Table \ref{sec:rel_work}).

This paper (ID 12 in Table \ref{tab:rel_work}) addresses these limitations by evaluating the impact of two cryptoeconomic incentives on human information-sharing behavior in a controlled experiment involving 132 participants over four days (Section \ref{sec:results}). The utilized token designs are illustrated (Section \ref{sec:exp_treatment}) and the (ethical) implications (Section \ref{sec:discussion}) of this work are discussed.

\section{Research methodology}
\label{sec:research_methodology}

The impact of two token incentives on human information-sharing behavior is investigated with an experimental methodology. The conducted experiment is explained below (Section \ref{sec:exp_meth}), followed by the measured variables (Section \ref{sec:variables_measures}), tested hypotheses (Section \ref{sec:hypotheses}) and the analysis methods (Section \ref{sec:analysis_meth}).

\subsection{Experiment}
\label{sec:exp_meth}
The experiment has been conducted by modifying the mixed-mode "living lab"~\cite{pournaras2022how} experimental methodology such that the randomized control trial is augmented with a 2x2 factorial pre-test/ post-test design that utilizes two token incentives (Figure \ref{fig:variables_methods}).

\begin{figure*}[th!]
\begin{floatrow}
\ffigbox{%
 \includegraphics[width=0.5\textwidth]{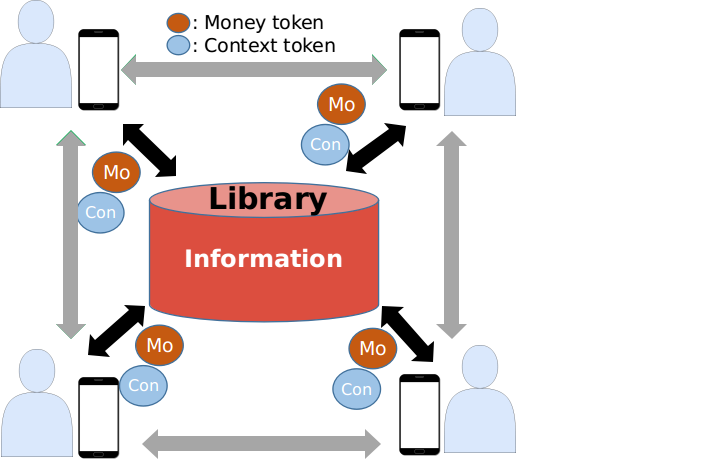}
}{%
   \caption{Experiment scenario: Participants share information with a library institution and obtain blockchain-based tokens in return. Tokens collected by other users can be discovered in interactions. }\label{fig:scenario}
}
\ffigbox{%
  \includegraphics[width=0.35\textwidth]{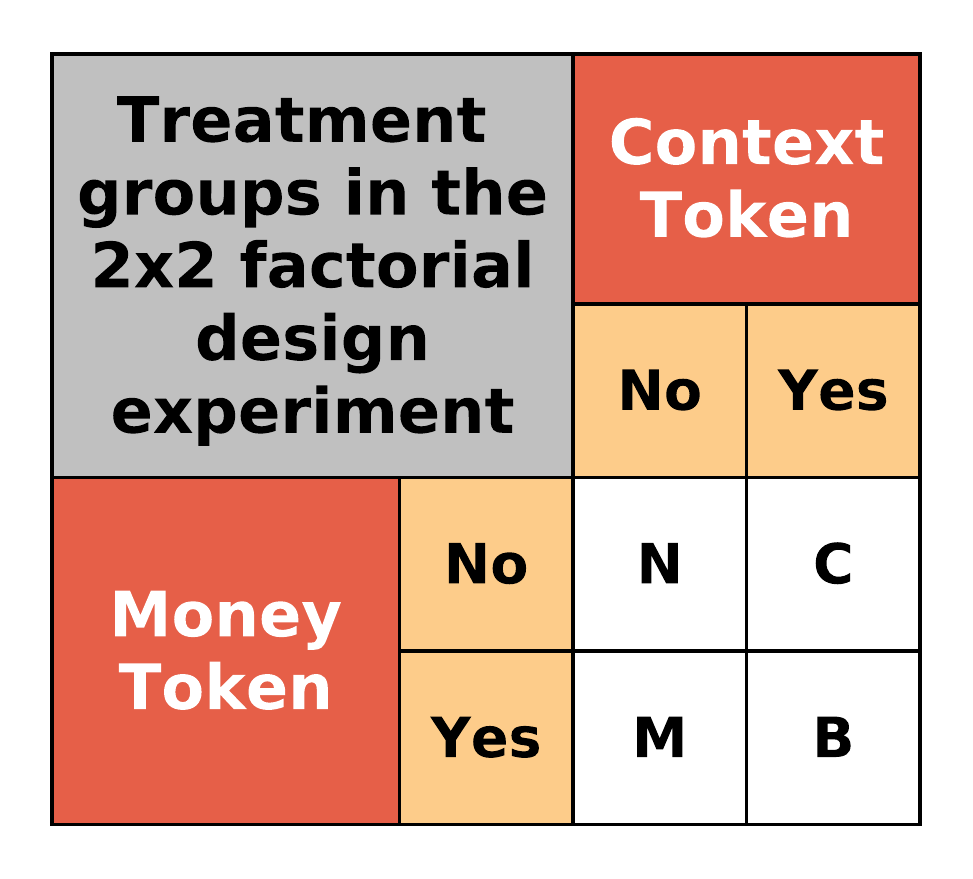}
}{%
   \caption{Treatment groups in the 2x2 factorial design experiment vary according to the received token incentive: N = no token incentives; C = context token incentive; M = money token incentive; B = both token incentives. }\label{fig:factorial_design}
}
\end{floatrow}
\end{figure*}

The experiment consists of three phases (Figure \ref{fig:variables_methods}): The entry and exit phase is facilitated by the ETH Decision Science Laboratory\footnote{\label{fn:descil}ETH Decision Science Laboratory provides infrastructures and services for researchers to perform human subject trials in the intersecting areas of Decision and Behavioral Sciences: https://www.descil.ethz.ch/ (last accessed: 2022-03-11)} (DeSciL) of ETH Zurich, using their infrastructure and staff. The core phase is facilitated by the research team and a blockchain-based Web 3.0 application (Section \ref{sec:exp_technical}). 
In the entry phase, participants provide their consent to the study and are instructed on the application of the software used in the core phase.
Before the core phase, the participants answer an entry survey consisting of demographic questions. The core phase of the experiment consists of four days in which participants utilize the software artifact to share information. On the second and third day of the core phase, in exchange for their shared information participants obtain token rewards.
In the exit phase, participants answer an exit survey and receive their financial compensation. 
The conducted experiment was granted an ethical approval by the Decision Science Laboratory (DeSciL) as well as the Ethics Commission of ETH Zurich. 

In the following, Section \ref{sec:exp_scenario} illustrates the real-world information-sharing scenario of the core phase, followed by Section \ref{sec:data_model}, where the model of the collected data is introduced. Section \ref{sec:exp_treatment} illustrates the applied incentives and the treatment groups. Then, Section \ref{sec:exp_technical} provides the technical specifications of the utilized software artifact. Finally, Section \ref{sec:exp_recruitment} provides an overview of the recruitment process and the compensation paid to the participants.

\subsubsection{Scenario}
\label{sec:exp_scenario}
The scenario of the experiment has been illustrated by \citet{ballandies2022improving}. A summary is given in the following and depicted in Figure \ref{fig:scenario}: Participants of the experiment share solicited information via their personal devices (e.g. laptop or mobile phone) with an organization, using a Web 3.0 application (Section \ref{sec:exp_technical}). In order to facilitate a realistic setup of the experiment and to comply with the anti-deception policy of DeSciL\footnotemark[\getrefnumber{fn:descil}], the shared information is received by a real-world library organization that has an interest in obtaining feedback from customers and unaware-customers\footnote{Unaware-customers are defined by the library as customers who are not aware that they are customers of the library. For instance, researchers accessing closed-access journals via their university credentials for which the library pays} of their services. Furthermore, in order to study user behavior in a realistic setting, participants can choose the time of feedback provision such that it is best integrated into their daily routines.
As an incentive for sharing information with the library organization, participants receive units of two types of blockchain-based tokens (Section \ref{sec:exp_treatment}). The amount of token units collected by other users and a subset of their shared information can be discovered in interactions.

The model of the shared information is illustrated in Section \ref{sec:data_model}.


\begin{figure*}[t!]
\begin{center}
\includegraphics[width=\textwidth]{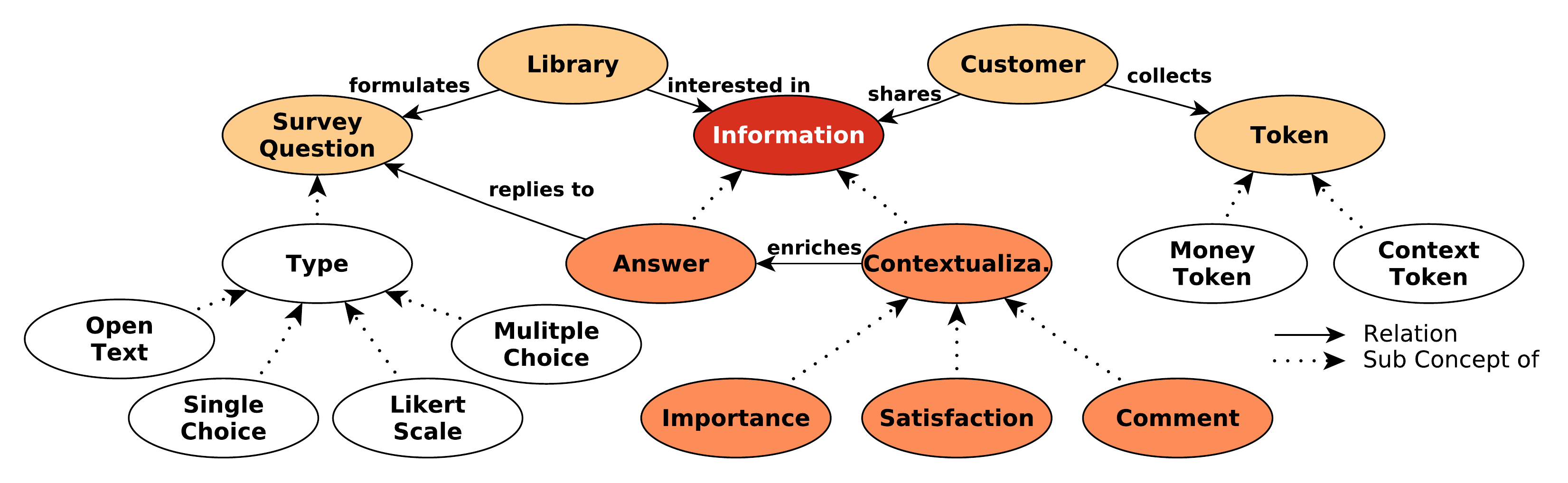}
\end{center}
\caption{Data model of the experiment that visualizes the stakeholders (library organization and its customers), collected information, survey questions, and token incentives.}\label{fig:data_model}
\end{figure*}

\subsubsection{Data Model}
\label{sec:data_model}

\begin{figure*}[h!]
\begin{floatrow}
\ffigbox{%
\includegraphics[width=0.5\textwidth]{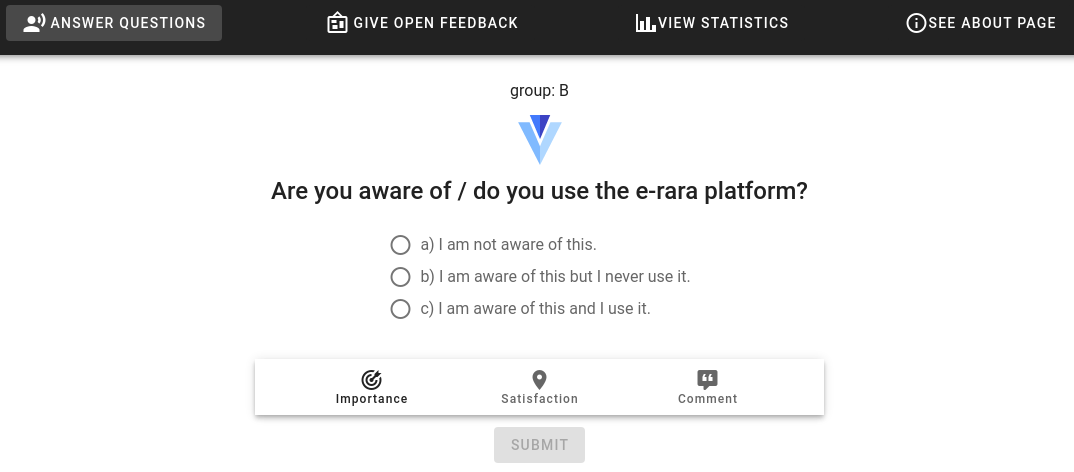}
}{
\caption{Answer view of the utilized software artifact: Users can answer questions posed by the library and contextualize it with three contextualization options (importance, satisfaction, and comment).} \label{fig:feed4org_answer}
}
\ffigbox{%
  \includegraphics[width=0.5\textwidth]{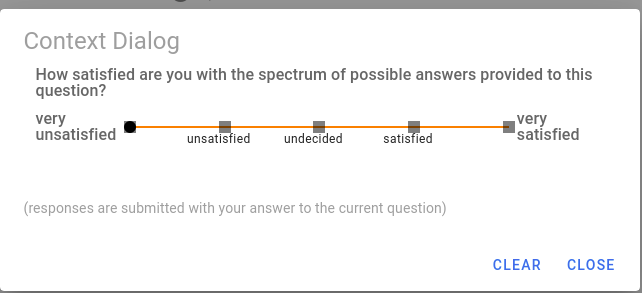}
}{%
   \caption{View of the satisfaction contextualization (Figure \ref{fig:feed4org_answer}). Users can specify how satisfied they are with the answer options provided for a question.}\label{fig:context}
}

\end{floatrow}
\end{figure*}

Figure \ref{fig:data_model} illustrates the ecosystem of the collected information as an ontology~\cite{ballandies2021mobile}. The library formulated $274$ survey questions, which they wanted to ask customers and unaware-customers of their services. These questions are of one of the following types: single-choice, multiple-choice, Likert scale, open text, or a combination of thereof.
In the experiment, participants take the role of customers and share information with the library in the form of answers to the given questions. Participants have the option to enrich their answer to a question with three types of contextualizations (Figure \ref{fig:feed4org_answer}). They can state from their perspective how important the question is for the library to improve their services (Likert scale), how satisfied they are with the answer options to the question (Likert scale), and provide a comment (open text field). 
As an incentive to share information, participants obtain units of two types of cryptoeconomic incentives: \textit{Money token} and \textit{Context token}, which are illustrated in greater detail in Section \ref{sec:exp_treatment}.

\begin{figure*}[t!]
\begin{center}
\includegraphics[width=0.75\textwidth]{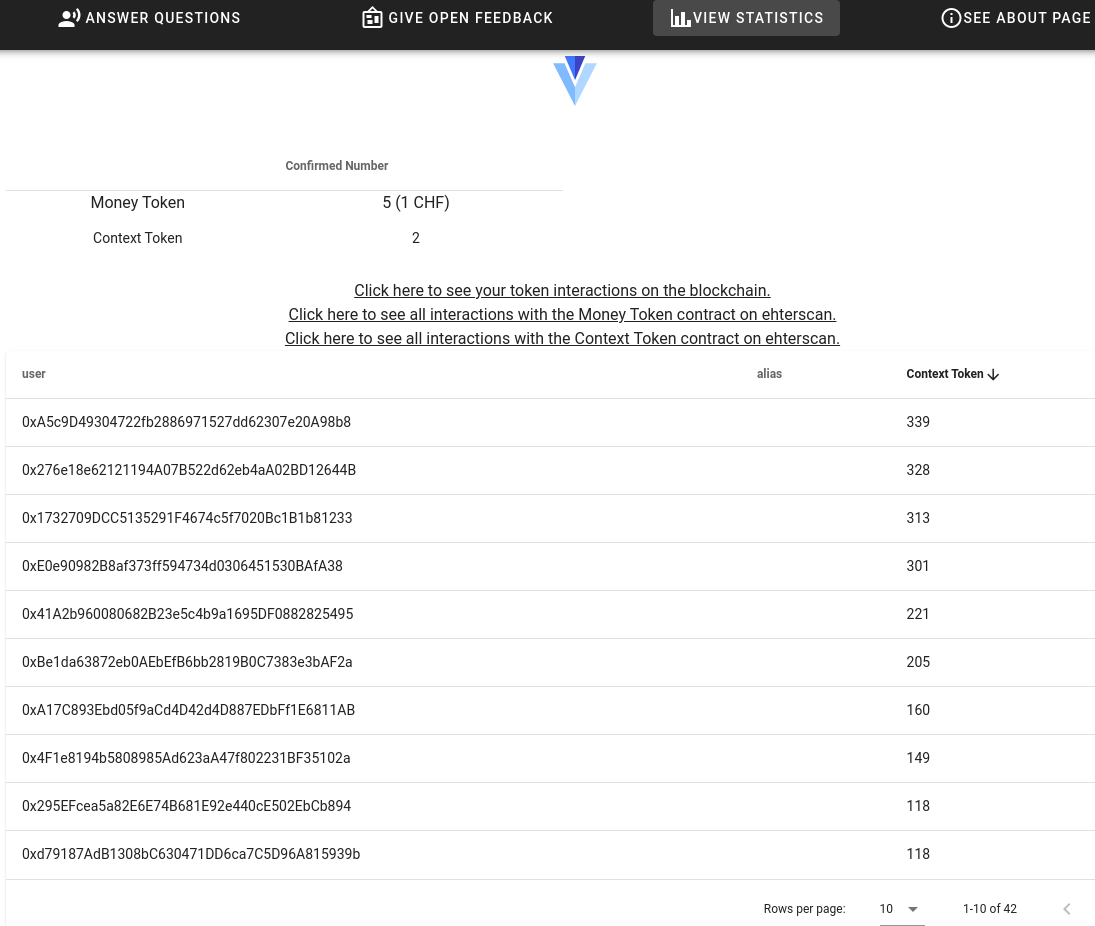}
\end{center}
\caption{Statistics view of the utilized software artifact. It depicts the amount of context token and money token units collected by a user (above). Moreover, the software artifact shows the leaderboard that compares users based on the collected context token units (below).}\label{fig:feed4org_statistics}
\end{figure*}

\subsubsection{Incentives and treatment groups}
\label{sec:exp_treatment}
Two types of cryptoeconomic incentives are utilized in this paper: The money token is a stable coin~\cite{pernice2019monetary} that resembles the Swiss fiat currency. It is a capped, pre-mined, transferable, and non-burnable ERC-20 token whose units are pegged to the Swiss franc at an exchange rate of 1:0.2 CHF. Users obtain a unit of this token whenever they provide an answer to a survey question (Figure \ref{fig:data_model}). 

The context token is a utility token~\cite{ballandies2021decrypting} and models reputation in the system: It is a ERC-20 token that is uncapped, transferable, burnable, and not pre-mined. A token unit is created whenever a contextualization (Figure~\ref{fig:data_model}) is performed in the system and is awarded to the user who provided that information. The amount of context token units collected is visible to others on a leaderboard during the experiment (Figure~\ref{fig:feed4org_statistics}) and thus constructs users' reputation, which functions as a source of value to this token. In particular, reputation is a widely adopted incentive mechanism that has been utilized to improve the quality of shared data~\cite{luo2019improving}. Additionally, the context token can be utilized to access a privileged service in the form of voting actions~\cite{ballandies2022improving}, which further provides value to the token~\cite{luo2019improving}.

Figure \ref{fig:factorial_design} illustrates the 2x2 factorial design of the study, whereby the two token types awarded to experiment participants are varied: Group N is the control group that receives no token incentives; Group C obtains the context token; Group M obtains the money token; Group B obtains both, the money and the context token.   

\subsubsection{Technical infrastructure}
\label{sec:exp_technical}
This research applies the customer feedback system developed by \citet{ballandies2022improving} to enable users to share information with a library organization and to receive two cryptoeconomic incentives (Section \ref{sec:exp_treatment}) in exchange. The software artifact is a Web 3.0 app that utilizes the Finance 4.0 infrastructure~\cite{dapp2021finance} and the Ethereum\footnote{https://ethereum.org/en/ (last accessed: 2022-03-21)} (ETH) blockchain. It enables the collection of solicited and unsolicited feedback from users of an organization. Figure \ref{fig:feed4org_answer} illustrates how users can provide solicited feedback by answering questions posed by an organization. This feedback can be contextualized by i) stating the importance of the question to improve the organization's service (bottom left in Figure \ref{fig:feed4org_answer}), or ii) stating the satisfaction with the answer options to the question (bottom center in Figure \ref{fig:feed4org_answer}), or iii) providing further feedback via a comment field (bottom right in Figure \ref{fig:feed4org_answer}). Figure \ref{fig:context} depicts how users can contextualize an answer with their satisfaction regarding the answer options. Figure \ref{fig:feed4org_statistics} shows how reputation is facilitated in the system by comparing users based on the amount of collected context token units. Moreover, this view gives users an overview of their collected money and context token units.

\subsubsection{Recruitment, compensation, and ethical approval}
\label{sec:exp_recruitment}
The participants were recruited by the ETH Decision Science Laboratory\footnotemark[\getrefnumber{fn:descil}] (DeSciL), who, following their protocols and ethical standards, were guaranteed fair compensation, and information regarding participants' identity was separated from the experiment data, thereby enabling anonymity for the participants. 
150 participants were recruited, 132 of which completed the exit phase (88 \% completion rate), which is a reasonable number that balances resources (compensation/ infrastructure), rigor, and control of the experimental process~\cite{pournaras2022how}. In particular, the mixed-mode experimental process preserves the realism of the scenario by involving a real-world organization that obtains the shared information, while facilitating controlled experimental conditions that result in a  novel high-quality dataset to allow (causal) inferences about human behavior under cryptoeconomic incentivization. 

Participants were recruited from the full UAST\footnote{https://www.uast.uzh.ch/ (last accessed: 2021-12-01)} pool (no criterion was applied), which mainly consists of students and researchers of ETH Zurich and the University of Zurich, and thus is subject to sampling biases when making inferences about the behavior of the general population. Nevertheless, as these are exactly the customers and unaware-customers of the real-world library organization around which the use case of information collection in this experiment was constructed, the participants' profiles match well to the experimental scenario. Consequently, the findings may be transferable to similar scenarios, where customers share information with an organization. 
Four recruitment sessions were performed within the period from May 17, 2021 to June 11, 2021.

The DeSciL requires the fair minimum and avarage compensation of experiment participants. This is satisfied by compensating each participant $i$ of a treatment group ($N,C,M,B$ in Figure \ref{fig:factorial_design}) Swiss francs via one of the following payout formulas $p$:

\begin{equation}
    \label{eq:payout}
    \begin{aligned}
        p(M_i) &= \text{min}\big(60 \text{ CHF}, \text{MT}(M_i) \times 0.2\text{ CHF}\big) \\ 
        p(B_i) &= \text{min}\big(60 \text{ CHF}, \text{MT}(B_i) \times 0.2\text{ CHF}\big) \\
        p(N_i) &= \text{max}\big(20 \text{ CHF}, \frac{T-P}{N(C) + N(B)}\big)\\
        p(C_i) &= \text{max}\big(20 \text{ CHF}, \frac{T-P}{N(C) + N(B)}\big)\\
    \end{aligned}
\end{equation}
where,
\begin{equation}
    \begin{aligned}
    \text{MT}(i) &: \text{amount of collected money token units of participant $i$} ,\\
    T &= N \times 40 \text{ CHF}; \text{total available payout} \\
    P &= \sum_i^{N(M)}p(M_i) + \sum_i^{N(B)}p(B_i);  \text{total received payout} \\ & \text{by groups M and B}\\
    N &:  \text{number of participants}\\
    N(j) &: \text{number of participants in treatment group $j$}
    \end{aligned}
\end{equation}
This results in a minimum compensation of 20 CHF and an average compensation of 40 CHF (0.5 CHF/min) for the participants.
The payout for participants of treatment groups that received the money token (groups M and B in Figure \ref{fig:factorial_design}) depends on the amount of  money token units collected. This amount is multiplied by $0.20$ CHF and then awarded to the participants. The total payout is capped at 60 CHF per participant resulting in a maximum of 150 questions for which a user can be rewarded per day.  
The payout for participants of the other treatment groups (groups N and C in Figure \ref{fig:factorial_design}) depend on the payout of the treatment groups that receive the money token (groups M and B in Figure \ref{fig:factorial_design}), such that the average compensation over all experiment participants is 40~CHF.

\subsection{Variables and measures}
\label{sec:variables_measures}
\begin{figure*}[h!]
\begin{center}
\includegraphics[width=\textwidth]{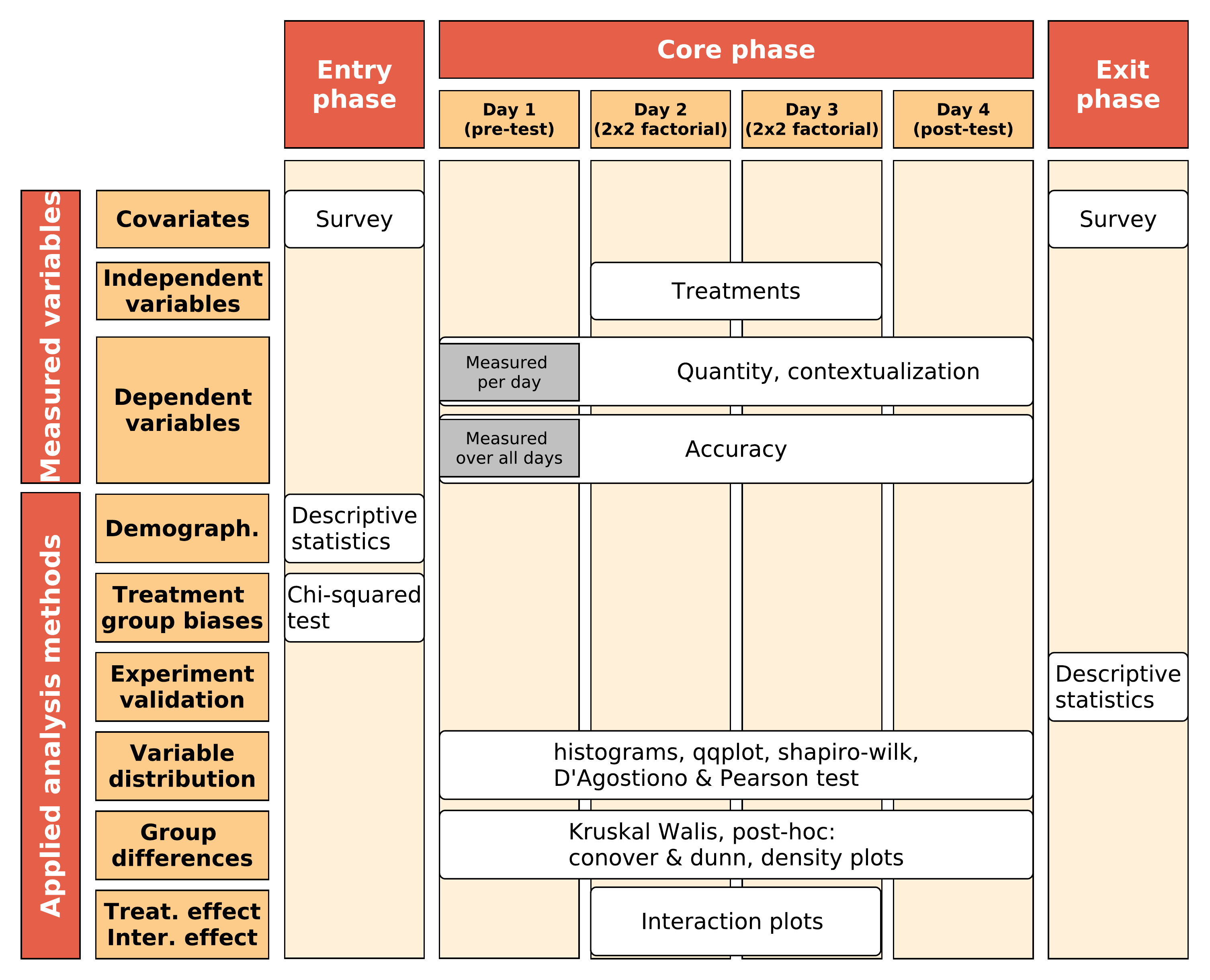}
\end{center}
\caption{Measured variables in the three phases of the experiment and the applied analysis methods.}\label{fig:variables_methods}
\end{figure*}

Figure \ref{fig:variables_methods} illustrates the measured variables of this paper in the three phases of the experiment. The participants answered demographic questions in the entry phase and another survey in the exit phase. 
Two extrinsic incentives, the money token, and the context token (Section \ref{sec:exp_treatment}) are manipulated as independent variables on the second and third day of the core phase.
Several dependent variables are measured each day: The quantity of shared information is measured by the number of replies to survey questions. Moreover, two quality characteristics are measured: i)
Contextualization is the number of contextualization actions performed by participants in response survey questions (via the bottom buttons in Figure \ref{fig:feed4org_answer} as shown in Section \ref{sec:exp_technical}). This is the amount of "metadata" that a user provides with an answer that contributes to the usability of information and is considered a quality dimension of information~\cite{cai2015challenges}. 
Further, ii) accuracy is a quality element of information that contributes to the reliability of information~\cite{cai2015challenges}. Applying the methodology of estimating choice variability~\cite{brus2021sources,polania2019efficient}, accuracy is operationalized in this paper as follows: With equal probability, survey questions are displayed more than once to participants. 
The average accuracy with which a participant answers a specific question is then calculated by taking the Jaccard similarity~\cite{jaccard1901bulletin} between the answers provided to that question. The final accuracy for a user is then obtained by taking the average similarity over all questions and days. 

\subsection{Hypotheses}
\label{sec:hypotheses}
The hypotheses of this paper test five assumptions regarding the utilized token incentives (Section \ref{sec:exp_treatment}) and treatment group assembly. They are formulated by connecting these assumptions to the introduced conceptual impact model~(Figure~\ref{fig:hypotheses_model}).
In the following, the five assumptions are first illustrated (Section \ref{sec:assumptions}) before the hypotheses are introduced (Section \ref{sec:hypotheses_formulation}).

\subsubsection{Assumptions}
\label{sec:assumptions}
\begin{assump}
\label{ass:money_token}
\textit{The money token (stable coin) is perceived as a monetary incentive and thus has a similar impact on the human information-sharing behavior as money. }
\end{assump}
The money token utilized in the experiment is a stable coin that has a fixed exchange rate with the Swiss franc (Section~\ref{sec:exp_treatment}) and thus resembles fiat money. The impact of fiat money on human behavior has been studied in information-sharing scenarios of related work (Section~\ref{sec:rel_work}). Due to this resemblance, it is hypothesized that the money token has a similar impact on human motivation and information-sharing behavior as monetary incentives. In particular, the money token impacts the extrinsic motivation positively and the intrinsic motivation negatively such that the quantity of shared information is increased and the quality is decreased, as illustrated in Figure~\ref{fig:hypotheses_model}.

\begin{assump}
\label{ass:context_reputation}
\textit{The context token impacts intrinsic motivation positively.} 
\end{assump}
The context token is a utility token that has reputation as its source of value (Section \ref{sec:exp_treatment}) and it is thus hypothesized that it is perceived as a competence-enhancing incentive~\cite{amabile1993motivational} that increases the intrinsic motivation of individuals. 

\begin{assump}
\label{ass:context_extrinsic}
\textit{The context token impacts extrinsic motivation positively.}
\end{assump}
Since the context token shares some characteristics with money (it is transferable and collectible), it is hypothesized that it has a positive impact on extrinsic motivation, albeit to a lower extend when compared to the money token. 

\begin{assump}
\label{ass:no_interaction}
\textit{No interaction exists between the money and the context token.}
\end{assump}
It is assumed that no interaction between the money and context token exists when they are applied simultaneously. 

\begin{assump}
\label{ass:no_bias}
\textit{No bias exists in the assembly of the treatment groups.}
\end{assump}
It is assumed that each treatment group consists of a similar participant structure.

These assumptions are the basis for the hypotheses that are formulated in the following.

\subsubsection{Hyptoheses formulation}
\label{sec:hypotheses_formulation}
In order to formulate the hypotheses, the assumptions (Section \ref{sec:assumptions}) are linked to the conceptual impact model (Figure \ref{fig:hypotheses_model}):

Under the assumption of no biases in the assembly of the treatment groups (Assumption~\ref{ass:no_bias}), it is hypothesized on Day~1 of the experiment that no difference in behavior among the treatment groups measured in quantity of answers or contextualizations are observed, because no token incentives are applied on that day:

\begin{hyp}
\textit{Day 1: \\ quantity(M) $=$ quantity(C) $=$ quantity(N) $=$ quantity(B)} 
\end{hyp}

\begin{hyp}
\textit{Day 1: \\ context(C) $=$ context(N) $=$ context(M) $=$ context(B)} 
\end{hyp}

Due to the impact on extrinsic motivation (Assumptions~\ref{ass:money_token} and~\ref{ass:context_extrinsic}), it is hypothesized from the arguments above that the M group (money token incentives) shares a greater quantity of information during incentivization days when compared to to the C group (context token incentive), which in turn shares more information than the N group (control group). Moreover, under the assumption of no interactions (Assumption~\ref{ass:no_interaction}) and because both tokens contribute to extrinsic motivation (Assumptions \ref{ass:money_token} and \ref{ass:context_extrinsic}), it is hypothesized that the B group (both token incentives) shares more information than the M group. Thus for Days 2 and 3, when incentives are applied, the following hypothesis is posed:

\begin{hyp}
 \textit{Days 2 \& 3: \\ quantity(B) $>$ quantity(M) $>$ quantity(C) $>$ quantity(N)}
\end{hyp}

Also, it is hypothesized that because of the competence-enhancing effect of the context token that would increase the intrinsic motivation of individuals (Assumption \ref{ass:context_reputation}), the C group shares information with greater quality characteristics such as contextualization or accuracy when compared to the N group. Moreover, because of the negative impact of the money token on intrinsic motivation (Assumptions \ref{ass:money_token}), the M groups quality characteristics are hypothesized to be worse than those of the N group. Finally, because i) the context token offsets the negative impact of the money token on intrinsic motivation (Assumptions \ref{ass:money_token} and \ref{ass:context_reputation}), and ii) there is no interaction effect between the tokens (Assumption \ref{ass:no_interaction}), it is hypothesized that the B group shares information with equal quality when compared to the N group, but less than the C group.
Thus for Day 2 and 3, when incentives are applied, the following hypotheses are stated:

\begin{hyp}
 \textit{Days 2 \& 3:\\ context(C)   $>$ context(B) $=$ context(N) $>$   context(M)}
\end{hyp}

Since no incentives are applied on the fourth day of the experiment, only the intrinsic motivation of individuals affects the characteristics of shared information on that day (Figure \ref{fig:hypotheses_model}). 
Thus, because it is assumed that the money token decreased (Assumption \ref{ass:money_token}) while the context token incentive increased (Assumption \ref{ass:context_reputation}) the intrinsic motivation, it is hypothesized that for the quality characteristics the C group outperforms the N group, which in turn outperforms the M group. Moreover, the N group and B group share an equal number of contextualizations:

\begin{hyp}
 \textit{Day 4: \\ context(C) $>$ context(N) $=$ context(B) $>$  context(M)}
\end{hyp}

In contrast, because intrinsic motivation only plays a minor role for the quantity of shared information (Figure \ref{fig:hypotheses_model}), it is hypothesized that the number of answers given on Day 4 does not differ significantly between the groups:

\begin{hyp}
\textit{ Day 4: \\ quantity(M) $=$ quantity(C) $=$ quantity(N) $=$ quantity(B)}
\end{hyp}

\begin{table*}[]
\caption{Results of the chi-squared ($\chi^2$) test for the eight demographic questions and the treatment/~wave grouping per treatment group/ recruitment wave illustrating that no bias is identified in the construction of the groups or the recruitment waves.} \label{tab:treatment_bias}
\begin{tabular}{lllllll} \hline\\
\textbf{}   & \textbf{}                                                            & \multicolumn{2}{c}{\textbf{Treat. Group}} & \multicolumn{2}{c}{\textbf{Rec. Wave}} & \multicolumn{1}{c}{\textbf{}}  \\ 
\textbf{ID} & \textbf{Question}                                                    & \textit{T. stat.}    & \textit{p val.}    & \textit{T. stat.}   & \textit{p val.}  & \multicolumn{1}{c}{\textbf{n}} \\ \hline \\
1           & What is your gender?                                                 & 1.5                  & 0.69               & 2                   & 0.57             & 3                              \\
2           & How old are you?                                                     & 49.5                 & 0.42               & 54.7                & 0.23             & 48                             \\
3           & How long do you use a mobile smart phone?                            & 8.9                  & 0.45               & 5.5                 & 0.79             & 9                              \\
4           & How long do you use blockchain/ crypto apps?                         & 7.2                  & 0.84               & 27.6                & 0.01             & 12                             \\
5           & Which of the following groups do you belong to?                      & 32.3                 & 0.22               & 22.1                & 0.73             & 27                             \\
6           & What is your role or function at {[}..{]} your institution {[}..{]}? & 18.1                 & 0.26               & 7.8                 & 0.93             & 15                             \\
7           & What field/subject do you mainly work/study in?                    & 26.1                 & 0.35               & 25.6                & 0.37             & 24                             \\
8           & Do you use the res. and services provided by {[}..{]} library?       & 6                    & 0.11               & 0                   & 1                & 3                              \\
9           & Recruitment Wave                                                                 & 2.1                  & 0.99               & 396                 & 0                & 9                              \\
10          & Treatment Group                                                           & 396                  & 0                  & 2.1                 & 0.99             & 9                 \\ \hline    
\end{tabular}
\end{table*}

\begin{table}[]
\caption{p-values obtained from the normality test of dependent variables per day/ over all days and treatment group. p-values $\geq$ 0.05 indicate normal distributions (marked with an asterisk).} \label{tab:normality}
\begin{tabular}{clllll} \hline \\
\multicolumn{1}{l}{\textbf{Day}} & \textbf{Treatment} & \multicolumn{3}{c}{\textbf{p-value}}                       \\
            &                    & \textit{Answ.} & \textit{Cont.} & \textit{Accu.}  \\ \hline \\
\multirow{4}{*}{\textbf{1}}      & \textit{No}        & 0              & 0              & -                     \\
                                 & \textit{Context}   & 0              & 0.002          & -                     \\
                                 & \textit{Money}     & 0              & 0              & -                     \\
                                 & \textit{Both}      & 0              & 0              & -                     \\
\multirow{4}{*}{\textbf{2}}      & \textit{No}        & 0              & 0.222*         & -                     \\
                                 & \textit{Context}   & 0              & 0.001          & -                     \\
                                 & \textit{Money}     & 0              & 0              & -                     \\
                                 & \textit{Both}      & 0              & 0              & -                     \\
\multirow{4}{*}{\textbf{3}}      & \textit{No}        & 0.762*         & 0.05*          & -                    \\
                                 & \textit{Context}   & 0.251*         & 0.04           & -                     \\
                                 & \textit{Money}     & 0.001          & 0              & -                     \\
                                 & \textit{Both}      & 0              & 0              & -                     \\
\multirow{4}{*}{\textbf{4}}      & \textit{No}        & 0.003          & 0              & -                     \\
                                 & \textit{Context}   & 0.16*          & 0.01           & -                     \\
                                 & \textit{Money}     & 0              & 0              & -                     \\
                                 & \textit{Both}      & 0              & 0              & -                     \\
\multirow{4}{*}{\textbf{All}}    & \textit{No}        & 0.001          & 0.231*         & 0                     \\
                                 & \textit{Context}   & 0.217*         & 0.002          & 0                 \\
                                 & \textit{Money}     & 0.528*         & 0              & 0.005                 \\
                                 & \textit{Both}      & 0              & 0.007          & 0.026                \\ \hline
\end{tabular}
\end{table}

The accuracy is measured over all four days. Consequently, a hypothesis about the daily differences among the groups cannot be drawn. Accuracy is a quality characteristic (Section \ref{sec:variables_measures}). Quality has been found to be positively impacted by intrinsic motivation (Section \ref{sec:rel_work}). Thus, it is hypothesized that the averaged accuracy score of the C group is higher than that of N group, which in turn has a higher score than the M group. Moreover, because the context token offsets the negative impact of the money token on intrinsic motivation, the N and B groups have a similar accuracy:

\begin{hyp}
\textit{ All days: \\ accuracy(C) $>$ accuracy (N) $=$ accuracy(B) $>$ accuracy (M)}
\end{hyp}



\subsection{Analysis methods}
\label{sec:analysis_meth}
Figure \ref{fig:variables_methods} illustrates the methods that are applied to evaluate the hypotheses. 
The demographic information from the entry survey in the entry phase is utilized to illustrate the profiles of participants. Moreover, this information is applied in chi-squared ($\chi^2$) tests~\cite{lowry2014concepts} to validate that no treatment group biases are present. In particular, the test is employed to test the null hypothesis that no relationship exists on the demographic variables among the treatment groups. 
Survey responses from the exit survey are used to validate the experimental setup such as the rewards obtained by the participants. 
Histograms, qqplots, the Shapiro-Wilk test~\cite{shapiro1965analysis,razali2011power} and the D'Agostiono \& Pearson test~\cite{d1971omnibus,d1973tests} are used to investigate the distribution of the dependent variables.
In order to analyze treatment group differences, the Kruskal-Wallis one-way analysis of variance by ranks for independent samples (H-test) is utilized~\cite{kruskal1952use} and, for a post-hoc pairwise comparisons test of mean rank sums, the Conover-Iman~\cite{conover1979multiple} and the Dunn~\cite{dunn1964multiple} methods are applied. Furthermore, CDF plots are utilized to investigate differences in group behavior.
The treatment effect of the applied incentives and the interaction effect among the incentives are analyzed via interaction plots. 

\begin{figure*}%
    \centering
    \subfloat[\centering Day 1]{{\includegraphics[width=0.245\linewidth]{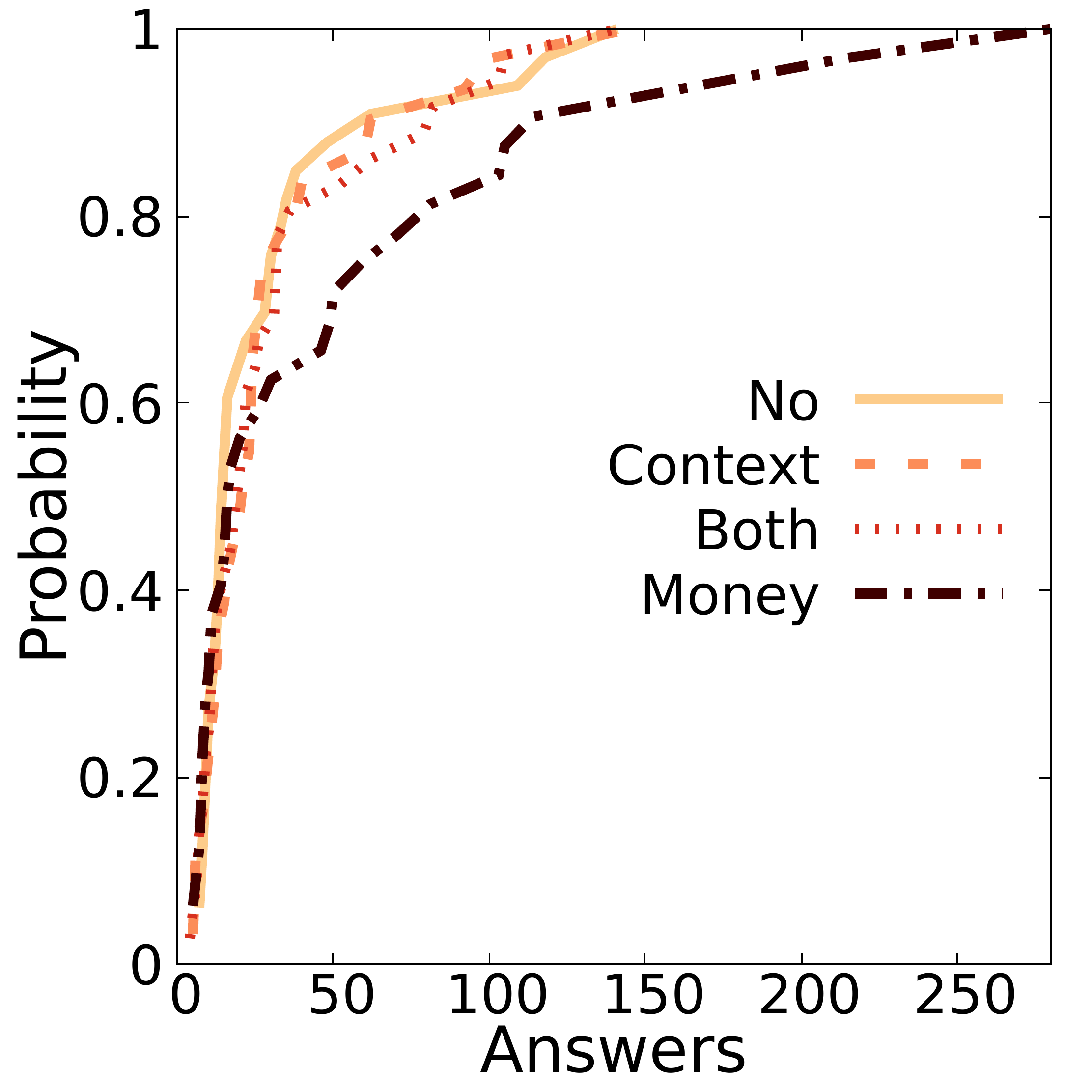} \label{fig:quant_day1_cdf}}}%
    \subfloat[\centering Day 2]{{\includegraphics[width=0.245\linewidth]{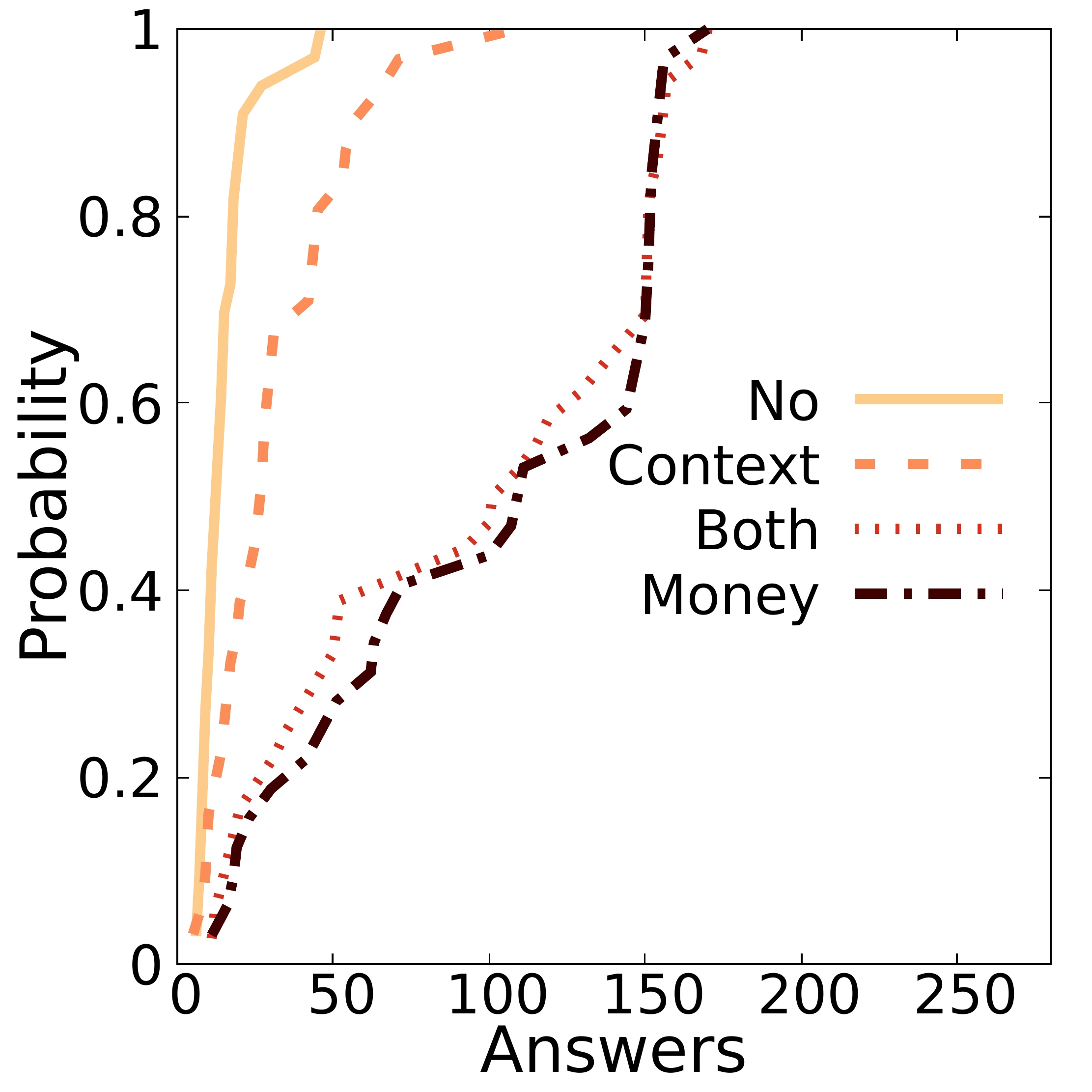} \label{fig:quant_day2_cdf}}}%
    \subfloat[\centering Day 3]{{\includegraphics[width=0.245\linewidth]{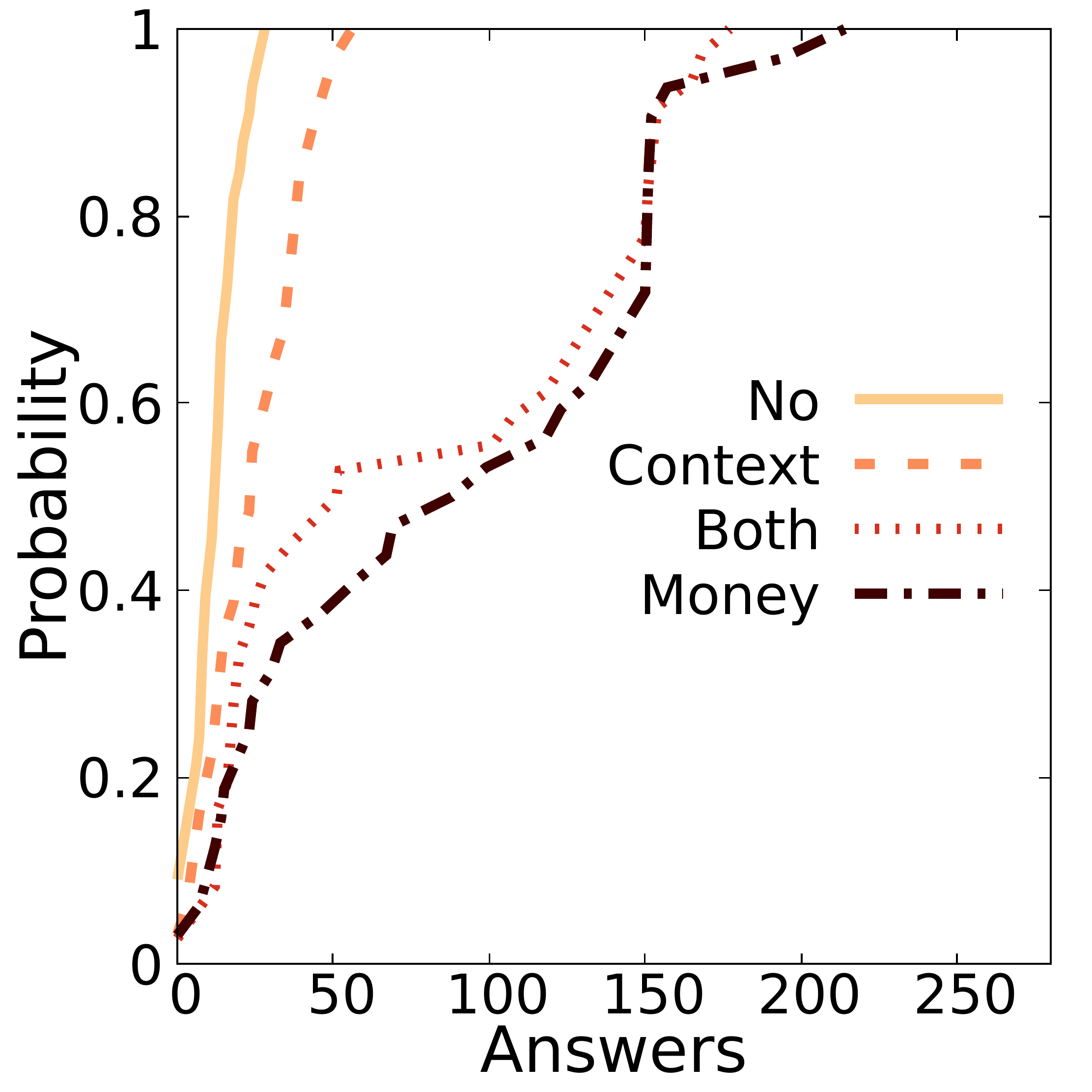} \label{fig:quant_day3_cdf}}}%
    \subfloat[\centering Day 4]{{\includegraphics[width=0.245\linewidth]{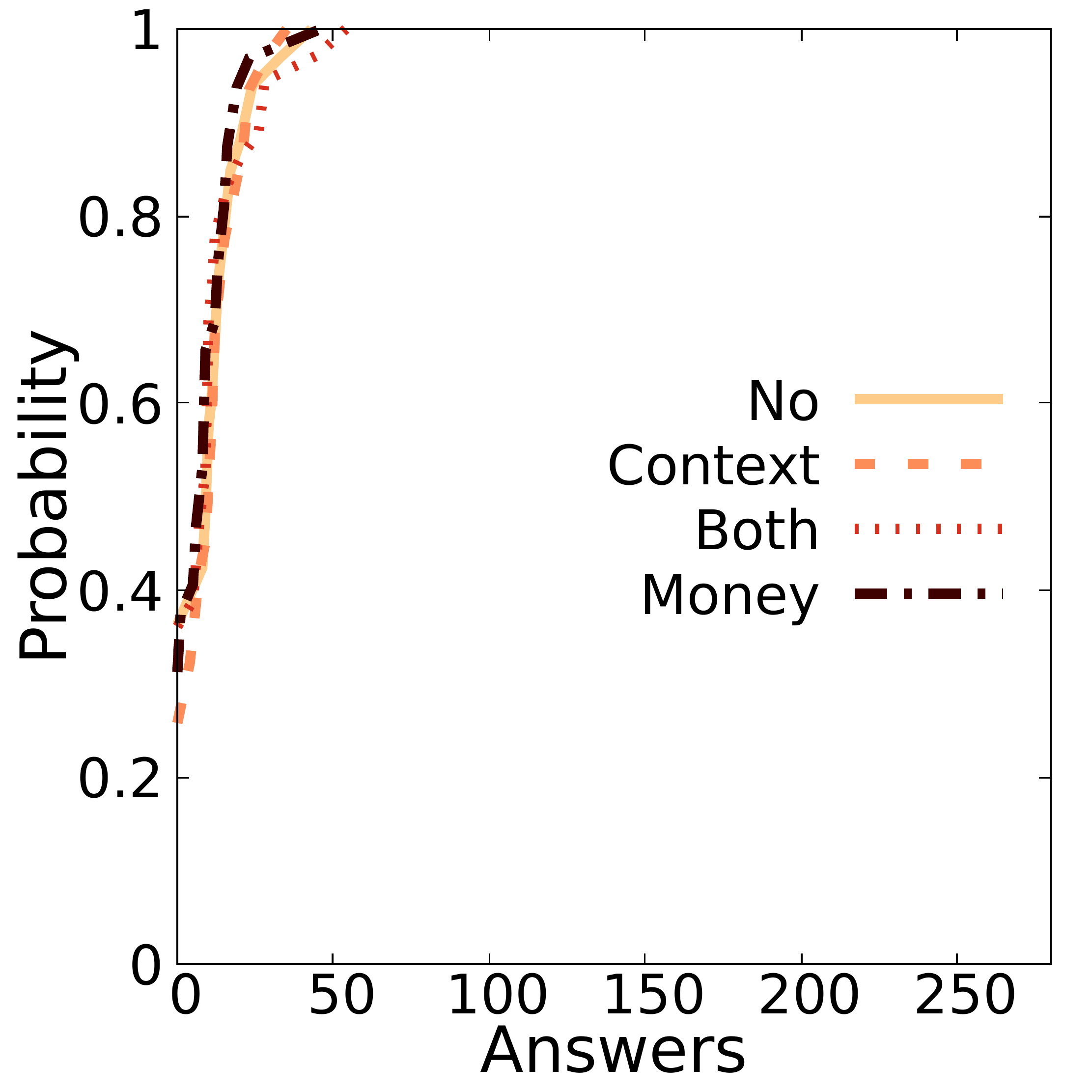} \label{fig:quant_day4_cdf}}}%
    \qquad
    \subfloat[\centering Day 1]{{\includegraphics[width=0.245\linewidth]{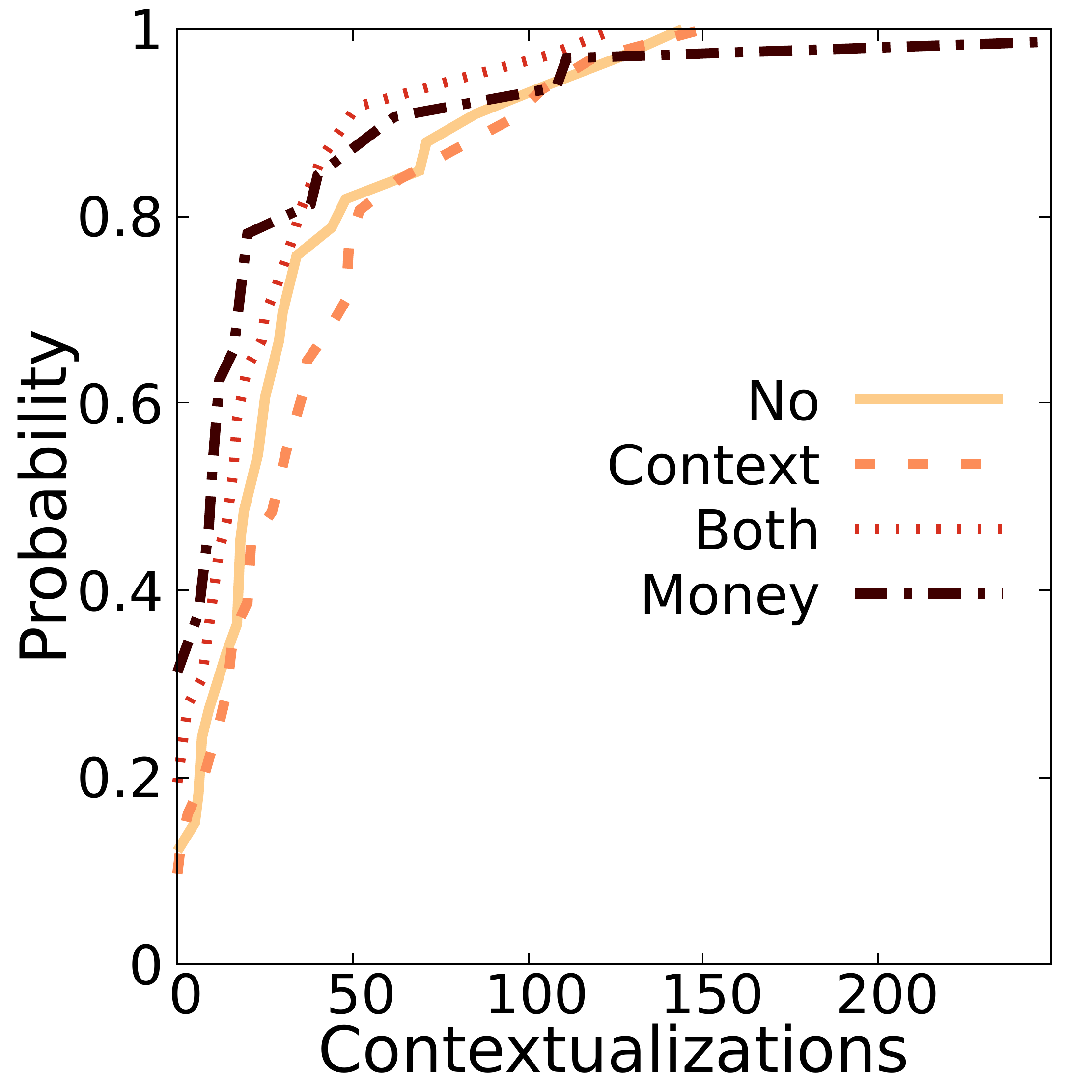}\label{fig:context_day1_cdf} } }%
    \subfloat[\centering Day 2]{{\includegraphics[width=0.245\linewidth]{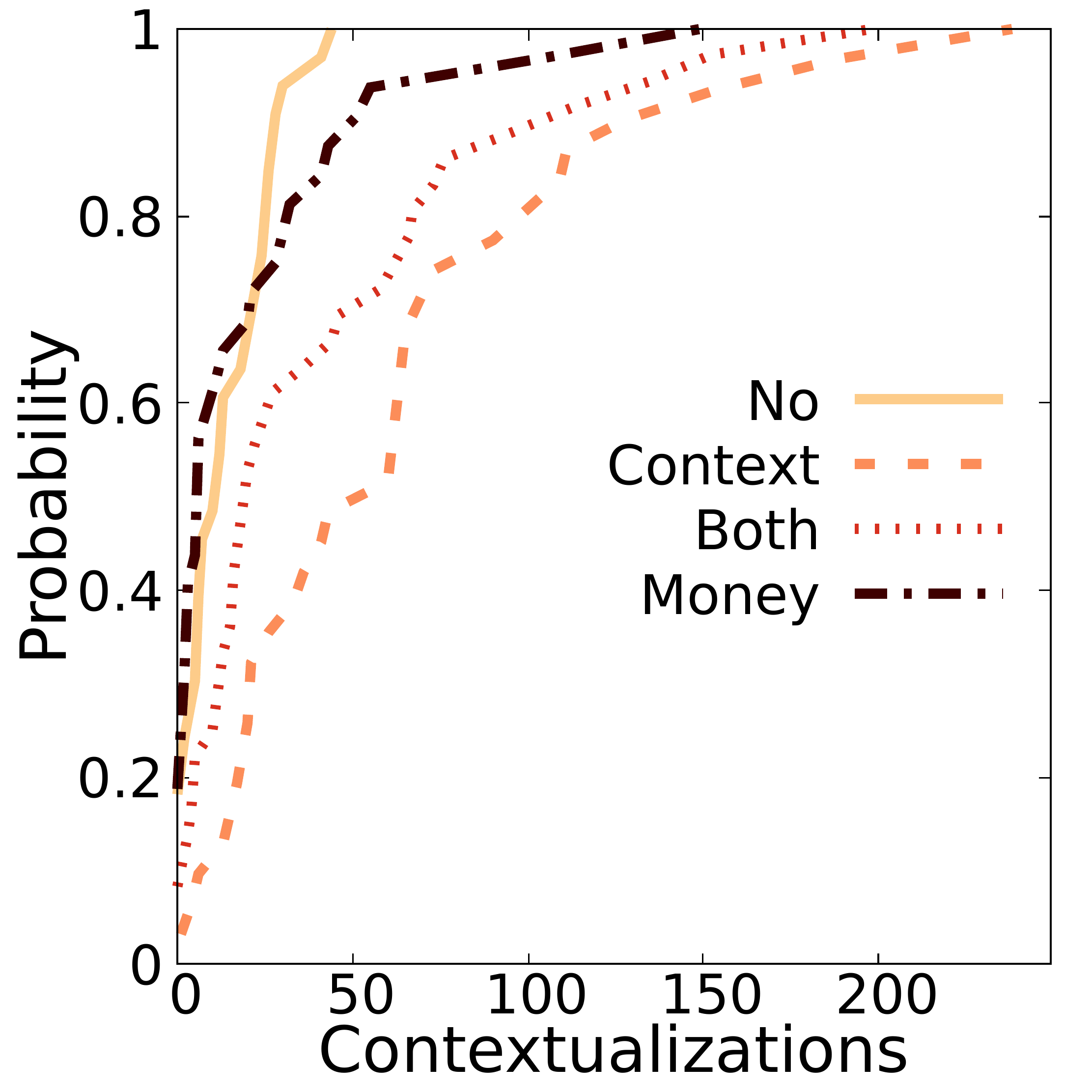}\label{fig:context_day2_cdf} }}%
    \subfloat[\centering Day 3]{{\includegraphics[width=0.245\linewidth]{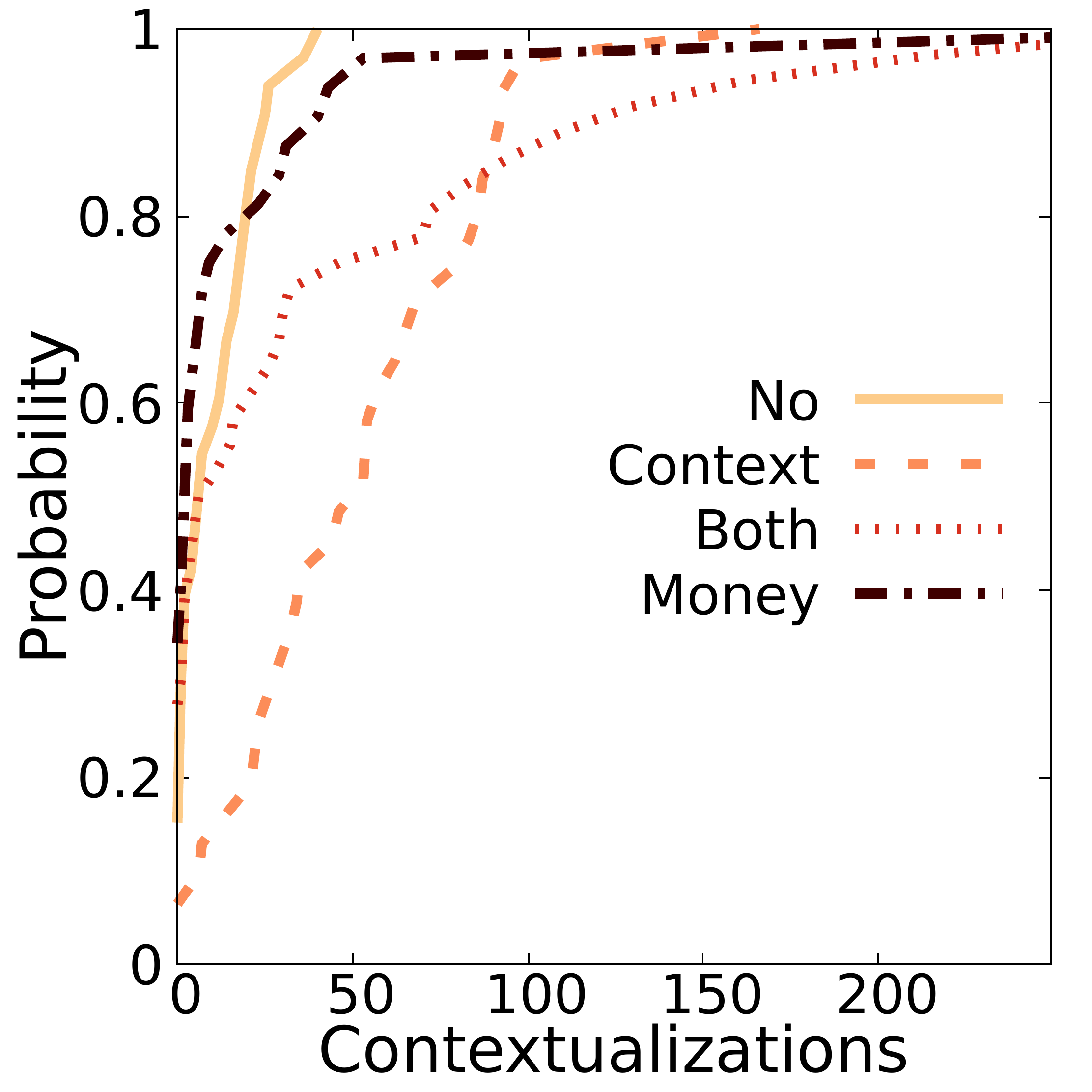}\label{fig:context_day3_cdf} }}%
    \subfloat[\centering Day 4]{{\includegraphics[width=0.245\linewidth]{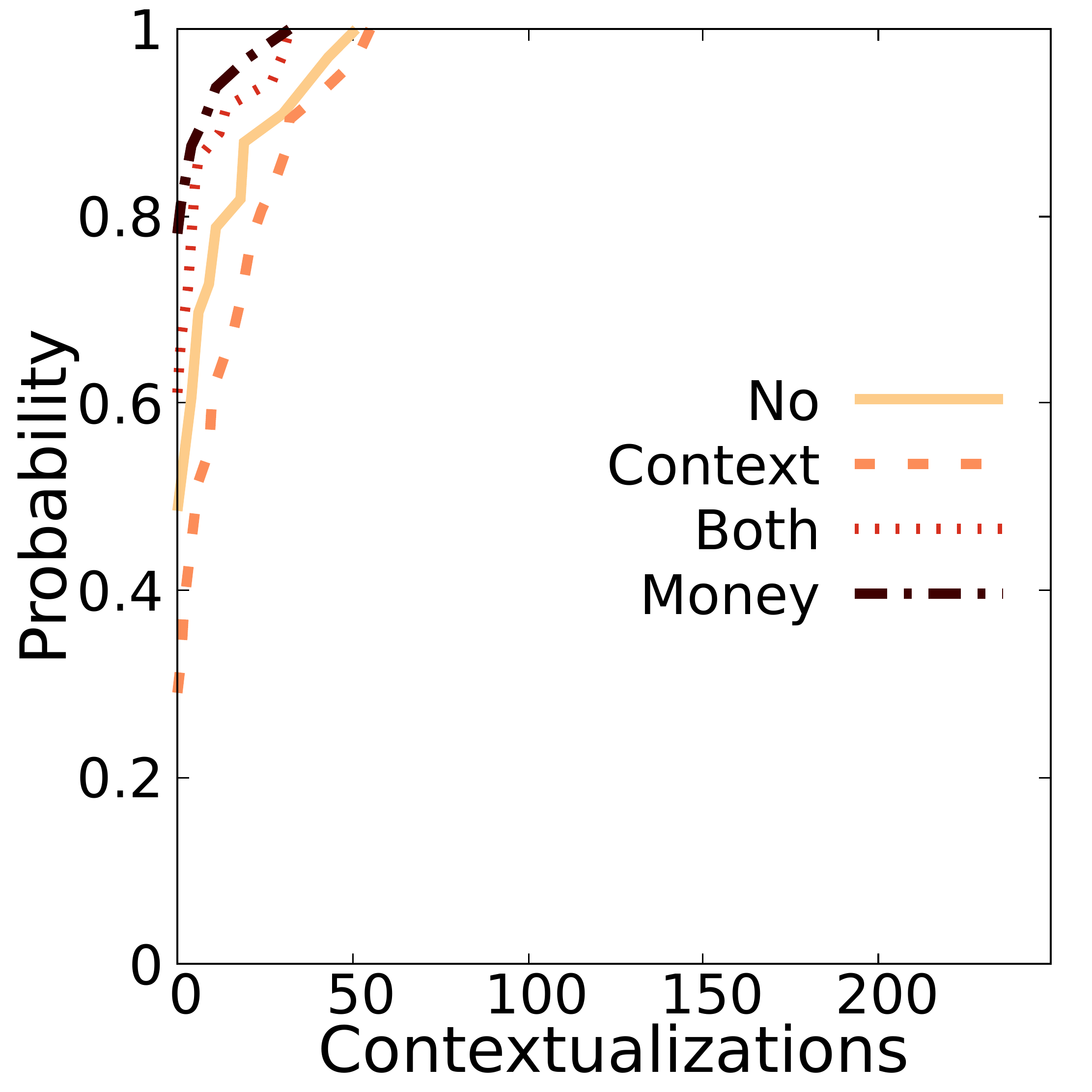} \label{fig:context_day4_cdf}}}%
    \qquad
    \subfloat[\centering All Days]{{\includegraphics[width=0.245\linewidth]{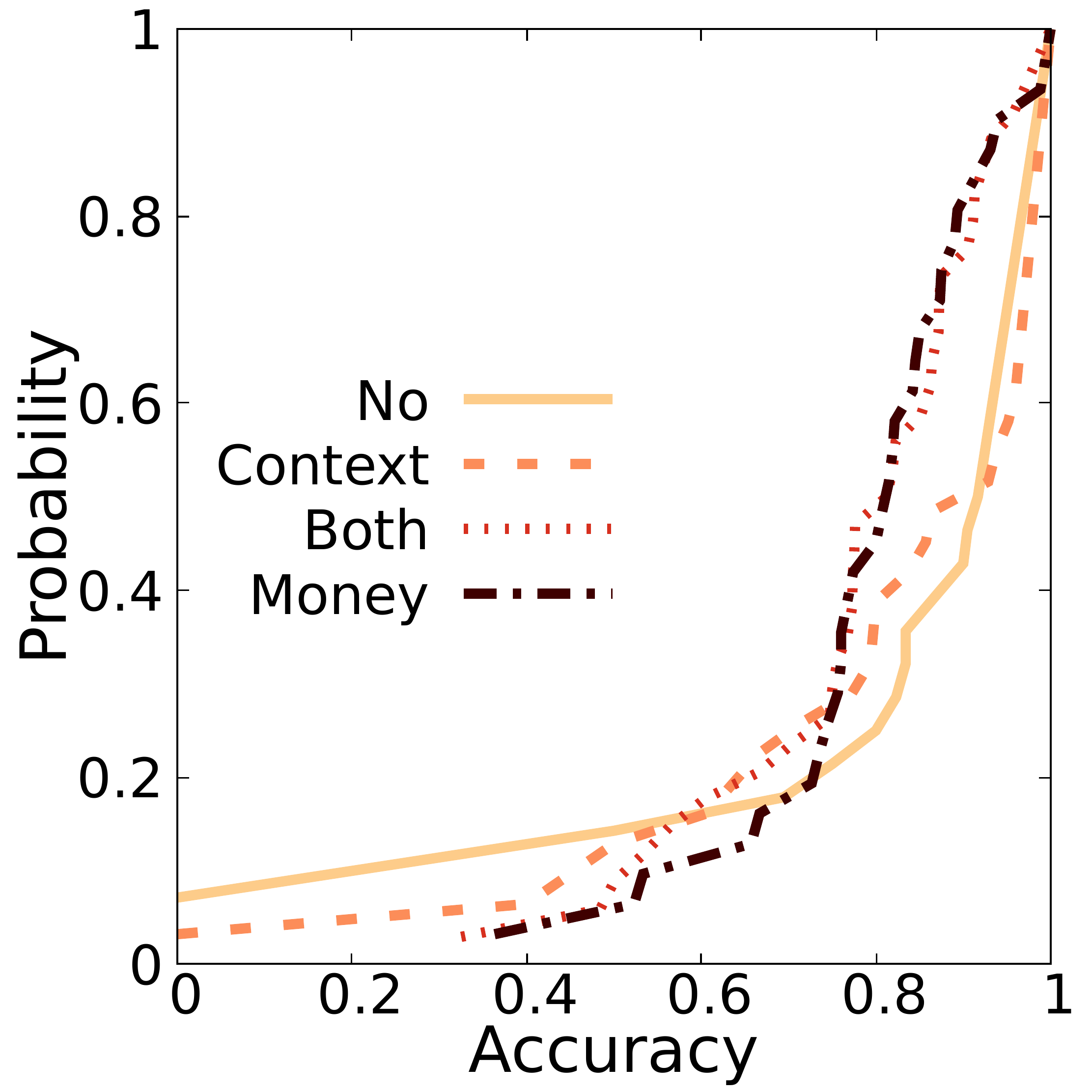} \label{fig:accurarcy_cdf} }}%
    \caption{Cumulative density plots for the treatments of the dependent variables over the four days/all days. The plot illustrates the cumulative percentage of users who reach an equal or lower value of the variable. }%
    \label{fig:cdf}%
\end{figure*}

\begin{figure*}%
    \centering
    \subfloat[\centering Day 1]{{\includegraphics[width=0.245\linewidth]{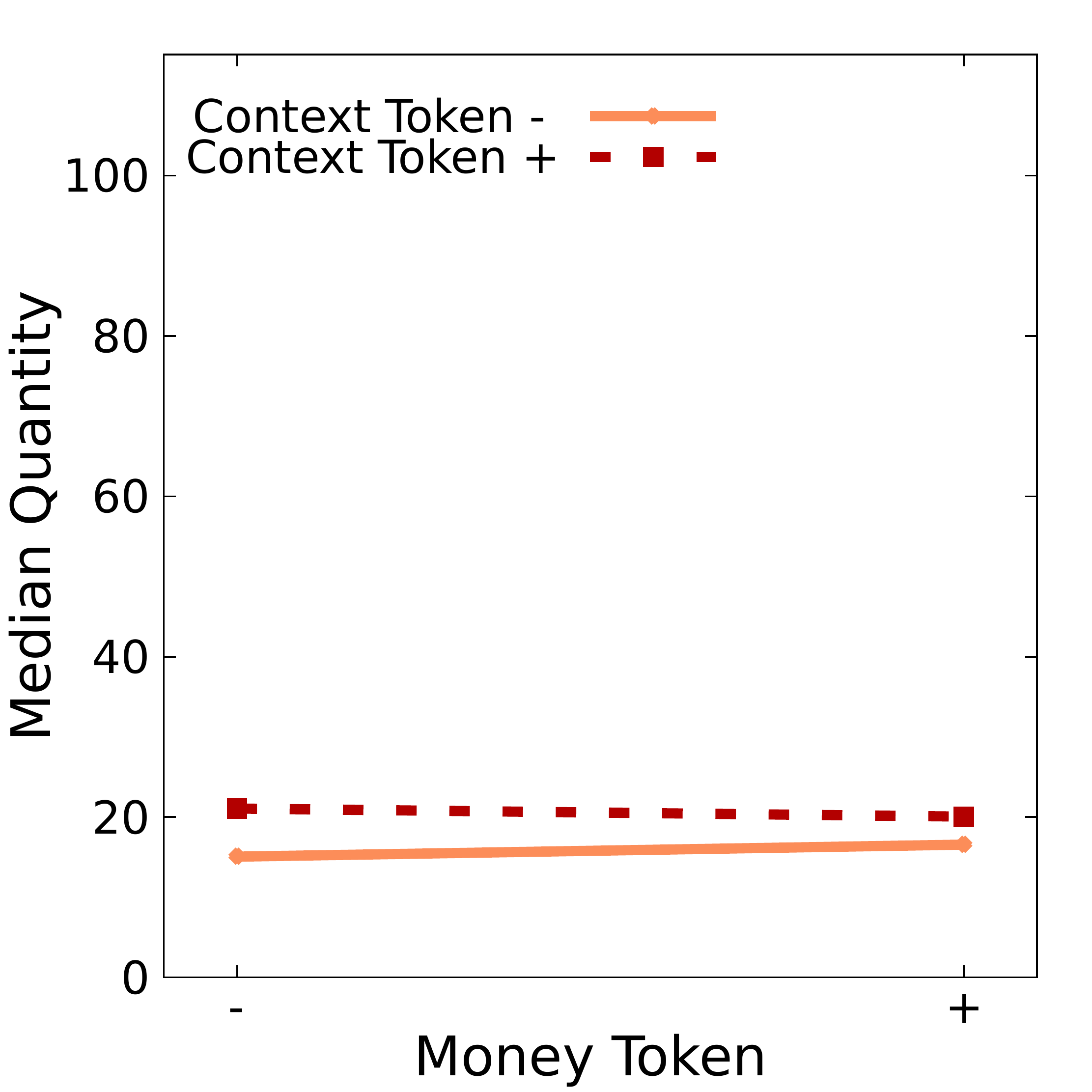} \label{fig:quantity_day1_interaction} }}%
    \subfloat[\centering Day 2]{{\includegraphics[width=0.245\linewidth]{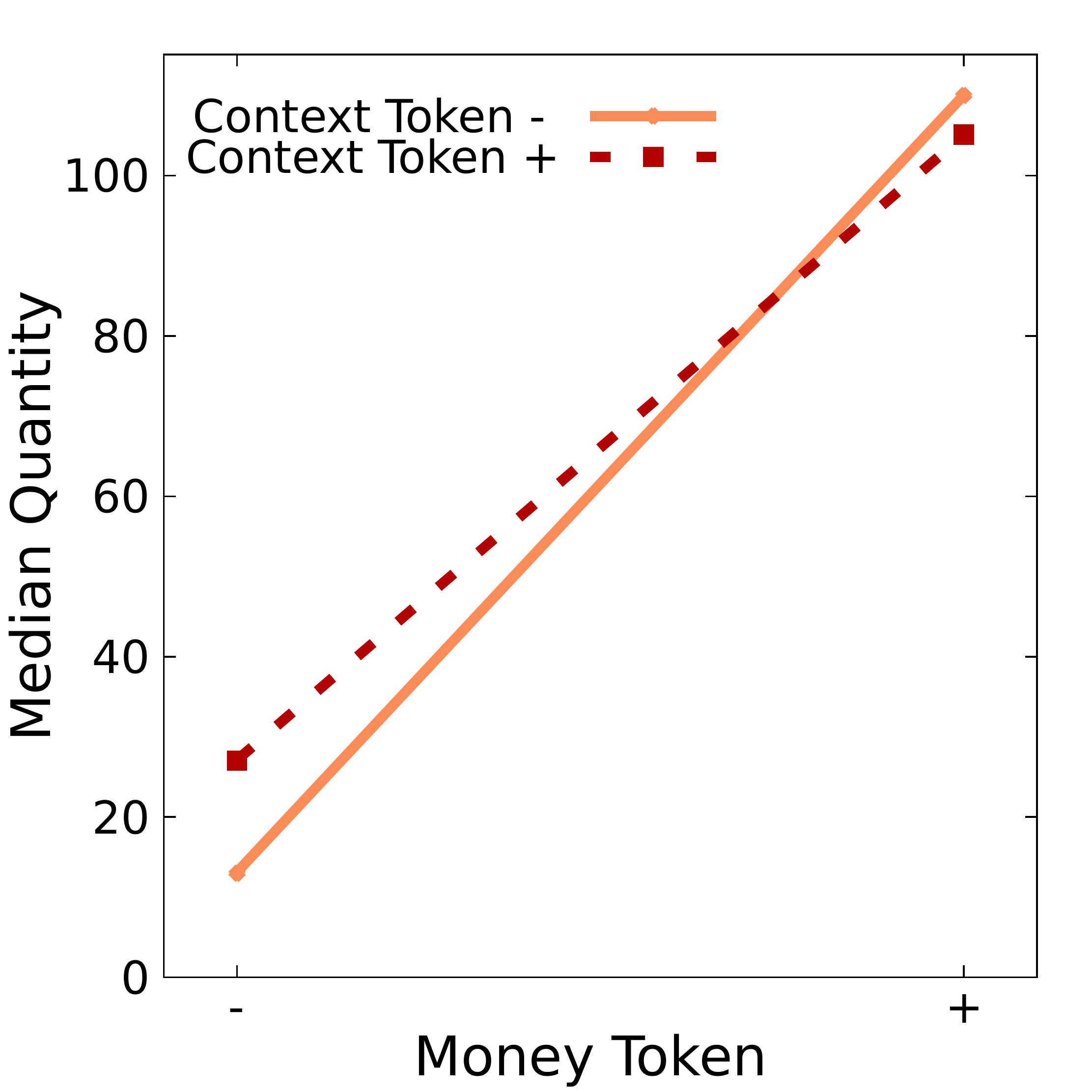} \label{fig:quantity_day2_interaction}}}%
    \subfloat[\centering Day 3]{{\includegraphics[width=0.245\linewidth]{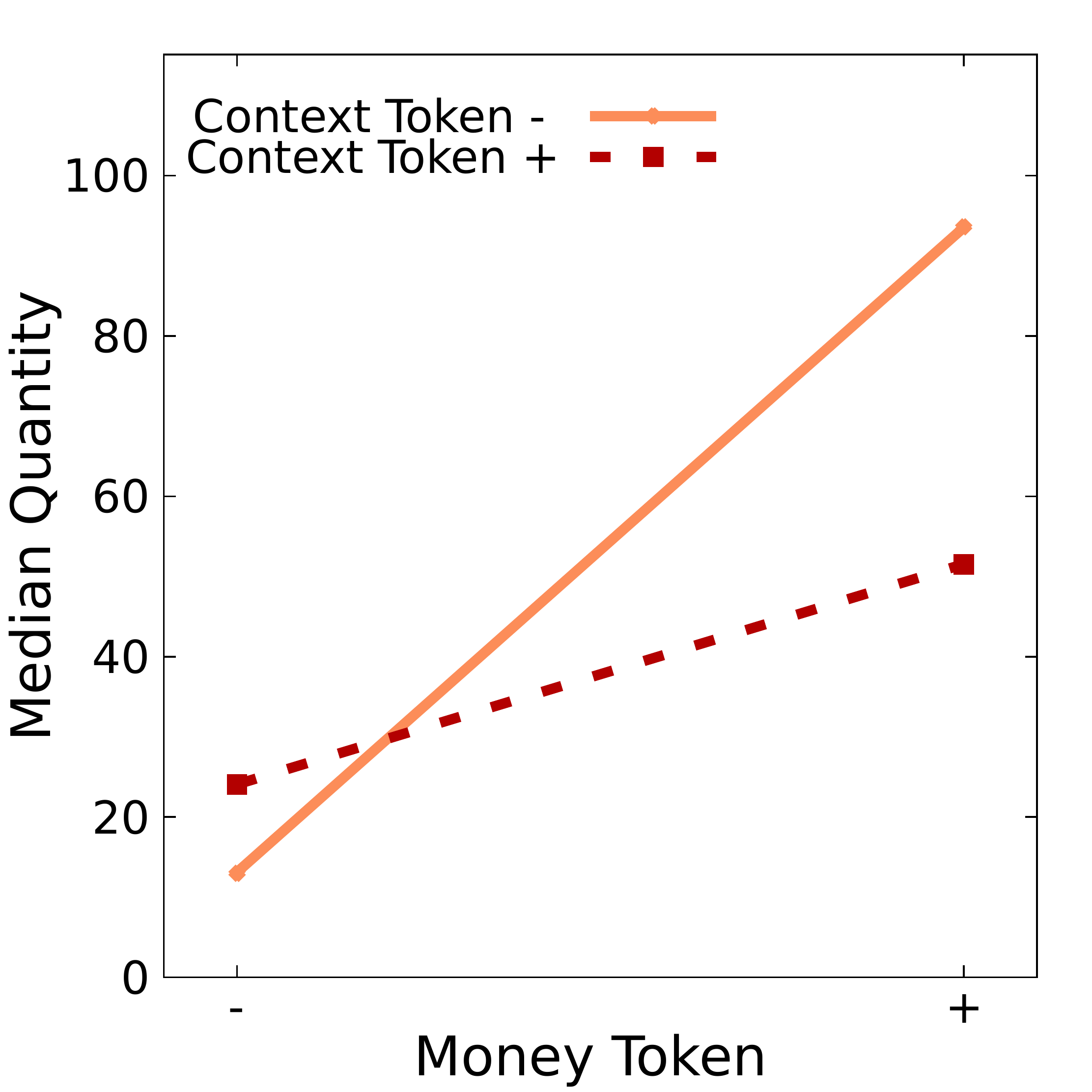} \label{fig:quantity_day3_interaction}}}%
    \subfloat[Day 4]{{\includegraphics[width=0.245\linewidth]{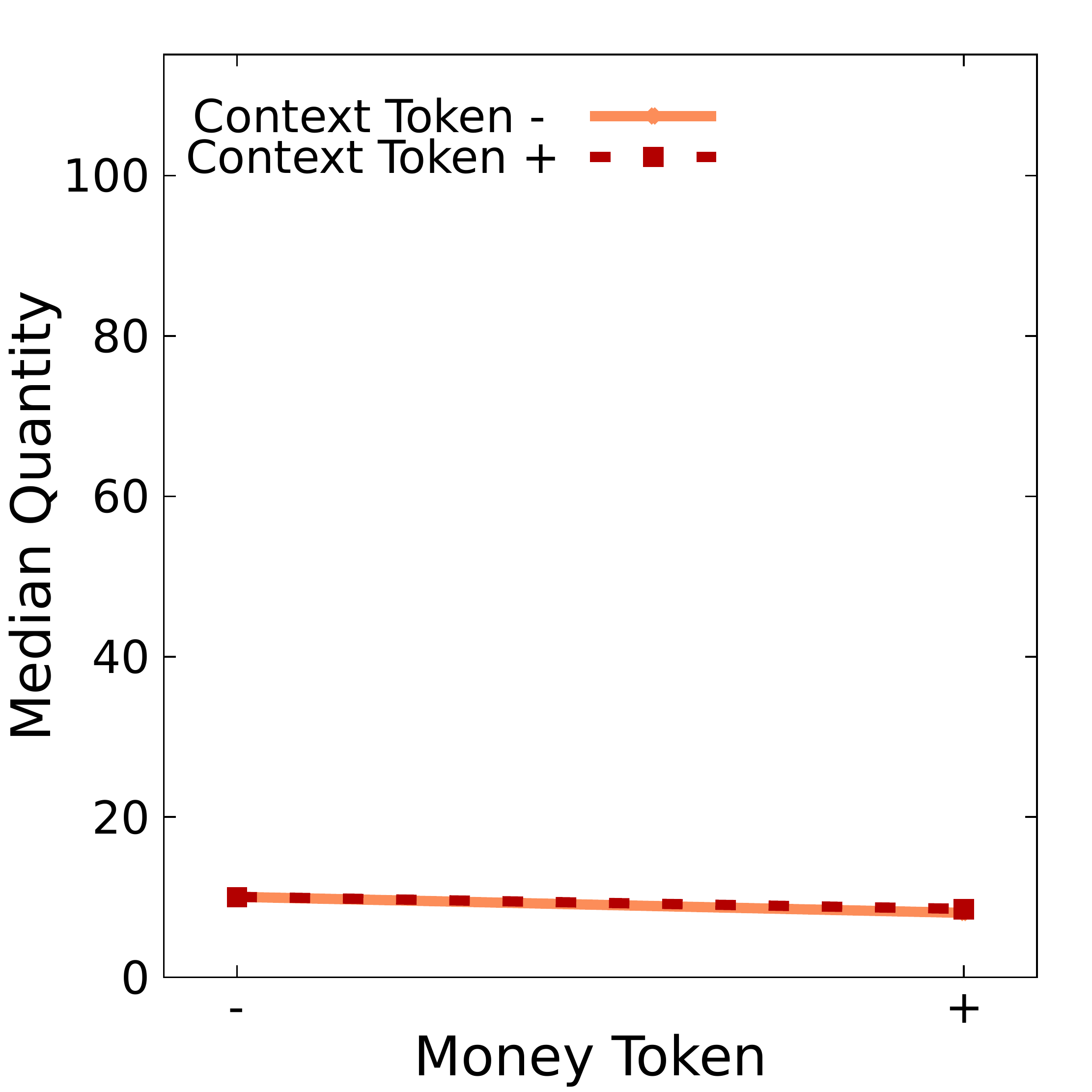} \label{fig:quantity_day4_interaction}}}%
    \qquad
    \subfloat[\centering Day 1]{{\includegraphics[width=0.245\linewidth]{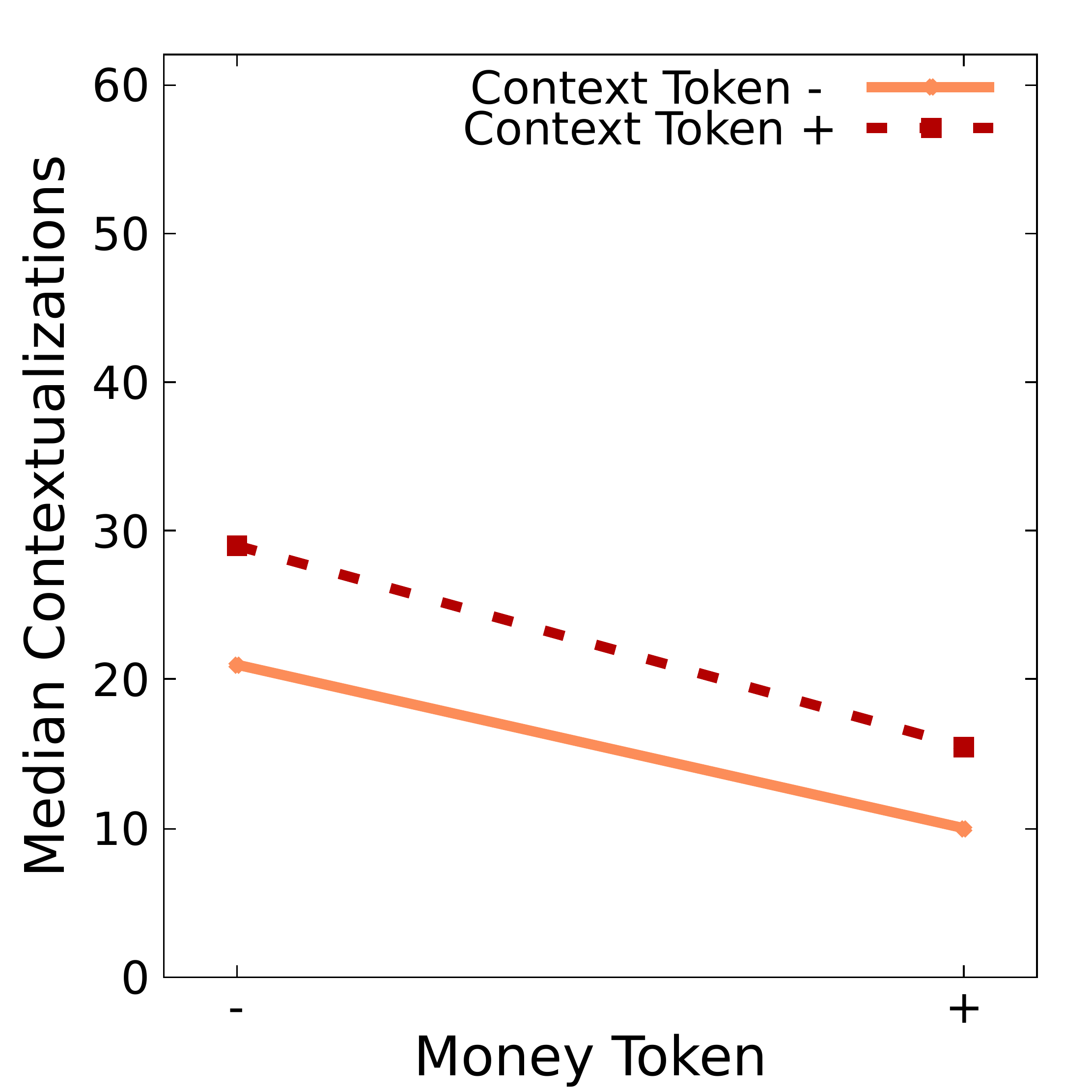}\label{fig:context_day1_interaction} } }%
    \subfloat[\centering Day 2]{{\includegraphics[width=0.245\linewidth]{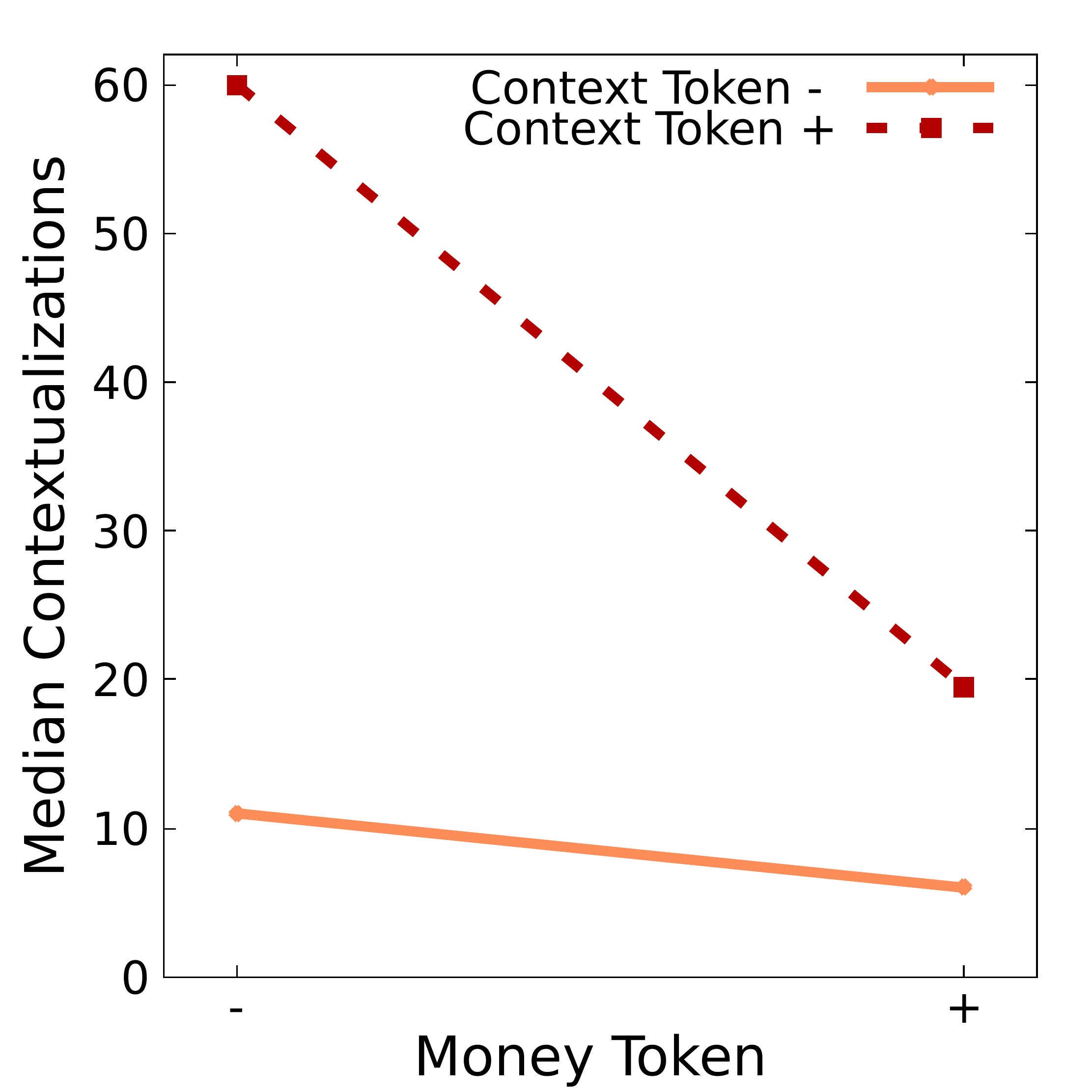}\label{fig:context_day2_interaction} }}%
    \subfloat[\centering Day 3]{{\includegraphics[width=0.245\linewidth]{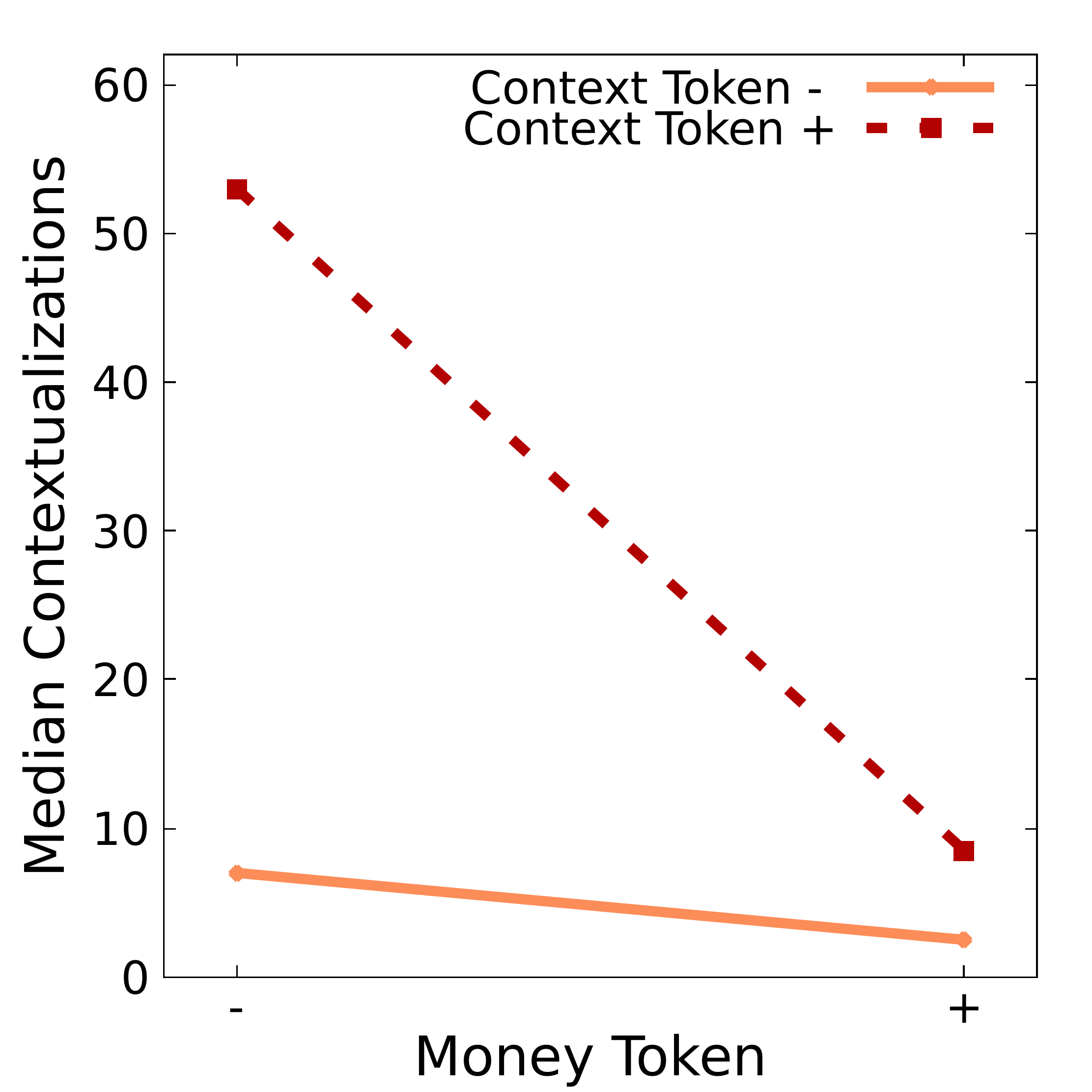}\label{fig:context_day3_interaction} }}%
    \subfloat[\centering Day  4]{{\includegraphics[width=0.245\linewidth]{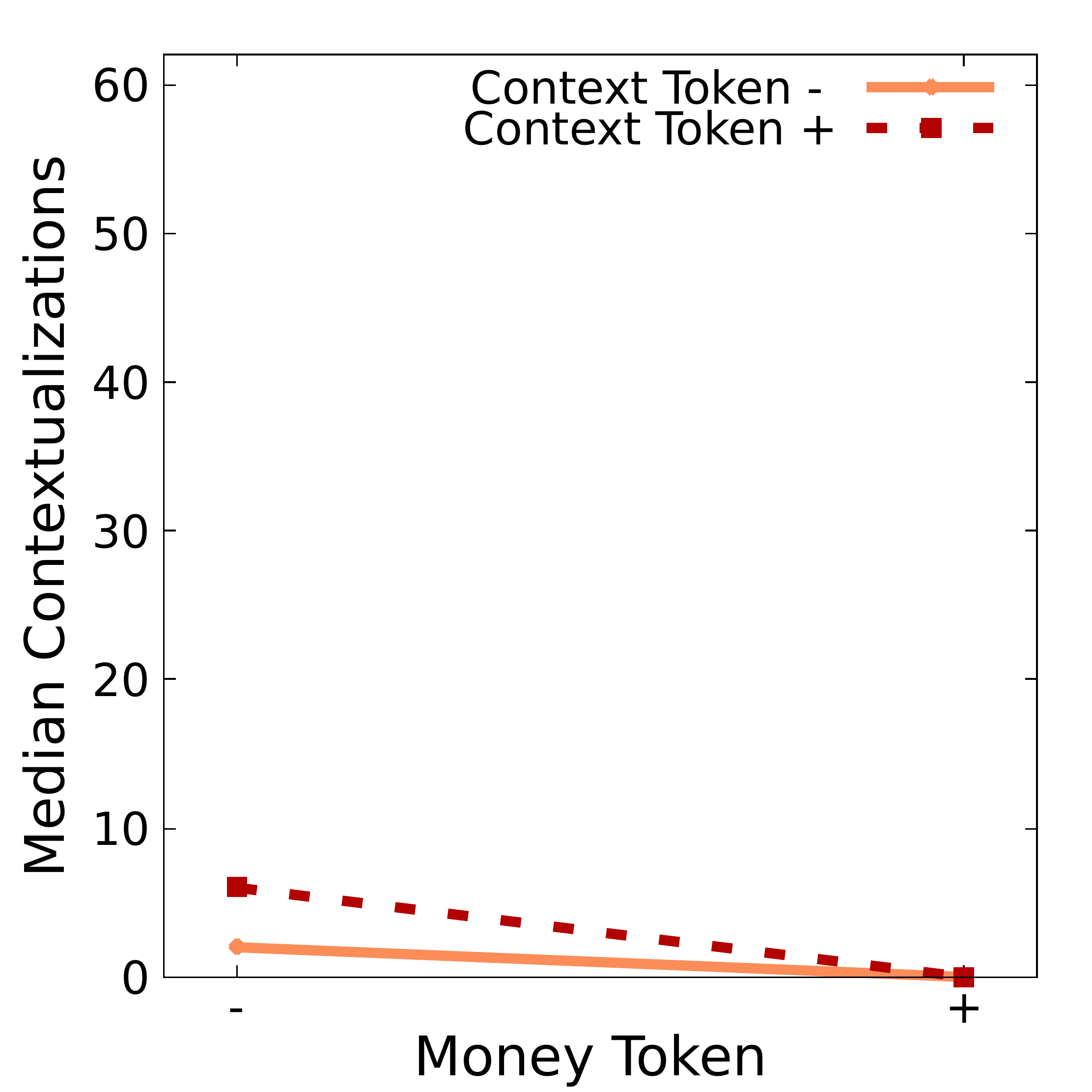}\label{fig:context_day4_interaction}}}%
    \qquad
    \subfloat[\centering All Days]{{\includegraphics[width=0.245\linewidth]{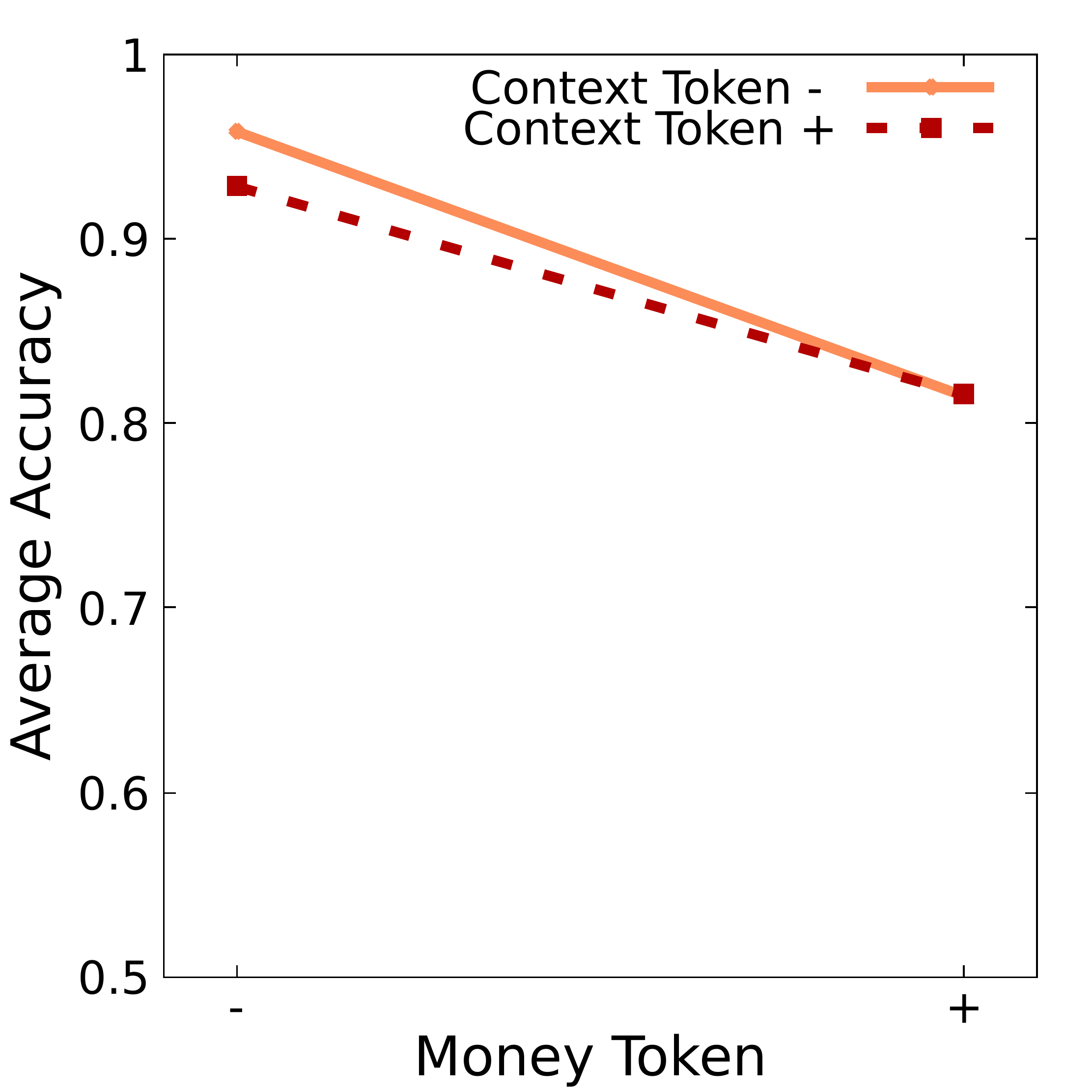} \label{fig:accuracy_interaction}}}%
    \caption{Interaction plots among the treatments for the dependent variables over the four days/ all days. The minus indicates that the token has not been applied, whereas the plus indicates an utilization of tokens. Thus, the dashed line connects the treatment groups that utilized the context token (left: treatment with context token; right: treatment with both tokens). The solid line connects the treatment groups that did not utilize the context token (left: treatment with no tokens (control group); right: treatment with money token).}%
    \label{fig:interaction_plots}%
\end{figure*}

\section{Results}
\label{sec:results}
\subsection{Demographics/ Profiles of the participants}
150 candidates were invited to participate in the study, $132$ of which completed all three phases (entry, core, and exit phase in Figure \ref{fig:variables_methods}). The average age of the participants was $23.2$ years. $62$ were male, $68$ were female and $2$ did not specify their gender. $36$ users had used blockchain/crypto apps before the experiment (50\% 1-6 month, 16.6\% 6-12 month, 16.6\% 1-2 years, 16.6 \% $>$ 2 years). $65$ participants were bachelor students, $56$ were master students, and 11 were "other". 54.5~\% of the participants stated that they were active users of the services of the library that functioned as a use case for the living lab experiment methodology.

\subsection{Treatment groups biases, experiment validation, and dependent variable distribution}

Table \ref{tab:treatment_bias} depicts the results of the chi-squared ($\chi^2$) test for the demographic questions and the treatment/wave per treatment~group/recruitment wave. Neither in the treatment group construction nor in the recruitment waves are biases identified. 

On average, the participants found the rewards fair (2.6/4)\footnote{Evaluated on a 5-point likert scale} and the onboarding materials useful (2.8/4). In particular, it has been identified in earlier work that learning how to utilize the web application is perceived as easy by the participants~\cite{ballandies2022improving}. Thus, the chosen compensation fulfilled the requirements of DeSciL\footnotemark[\getrefnumber{fn:descil}] and the chosen technology in the form of a blockchain-based Web application did not restrict users in participating in information sharing.

Table \ref{tab:normality} illustrates the distributions of the dependent variables utilized in the analysis. In the majority of cases, these variables are non-normally distributed, thus requiring the Kruskal-Wallis test that does not assume normally distributed variables to analyze the distribution~\cite{kruskal1952use}.

\subsection{Group differences and interaction effects}

\begin{table*}[] \caption{Findings from the Kruskal-Wallis and Conover-Iman posthoc analysis. Light green are those entries marked that accept the hypotheses stated in Section \ref{sec:hypotheses}. The following deviations can be explained by adjusting the assumptions of the hypotheses: i - Context token has a negligible effect on intrinsic motivation; ii - Extrinsic motivation has a considerable impact on contextualization; iii - Interaction effects between the tokens are present.} \label{tab:findings}
\begin{tabular}{lllccccccc} \hline \\
{\color[HTML]{4C4C4C} \textbf{ID}} & {\color[HTML]{4C4C4C} \textbf{Character.}} & {\color[HTML]{4C4C4C} \textbf{Hyp.}} & {\color[HTML]{4C4C4C} \textbf{Day}} & \multicolumn{6}{c}{{\color[HTML]{4C4C4C} \textbf{Finding}}}                                                                                                                                                                                                                                                                                         \\
{\color[HTML]{4C4C4C} }            & {\color[HTML]{4C4C4C} }                    & {\color[HTML]{4C4C4C} }              & {\color[HTML]{4C4C4C} }             & {\color[HTML]{4C4C4C} \textit{a}}                      & {\color[HTML]{4C4C4C} \textit{b}}                          & {\color[HTML]{4C4C4C} \textit{c}}                      & {\color[HTML]{4C4C4C} \textit{d}}                    & {\color[HTML]{4C4C4C} \textit{e}}                      & {\color[HTML]{4C4C4C} \textit{f}}                    \\ \hline \\
{\color[HTML]{4C4C4C} 1}           & {\color[HTML]{4C4C4C} Quantity}            & {\color[HTML]{4C4C4C} H1}            & {\color[HTML]{4C4C4C} 1}            & \cellcolor[HTML]{E5F5E0}{\color[HTML]{4C4C4C} $M =C$}  & \cellcolor[HTML]{E5F5E0}{\color[HTML]{4C4C4C} $C=N$}       & \cellcolor[HTML]{E5F5E0}{\color[HTML]{4C4C4C} $M=N$}   & \cellcolor[HTML]{E5F5E0}{\color[HTML]{4C4C4C} $B=M$} & \cellcolor[HTML]{E5F5E0}{\color[HTML]{4C4C4C} $B=C$}   & \cellcolor[HTML]{E5F5E0}{\color[HTML]{4C4C4C} $B=N$} \\
{\color[HTML]{4C4C4C} 2}           & {\color[HTML]{4C4C4C} Quantity}            & {\color[HTML]{4C4C4C} H3}            & {\color[HTML]{4C4C4C} 2}            & \cellcolor[HTML]{E5F5E0}{\color[HTML]{4C4C4C} $M > C$} & \cellcolor[HTML]{E5F5E0}{\color[HTML]{4C4C4C} $C>N$}       & \cellcolor[HTML]{E5F5E0}{\color[HTML]{4C4C4C} $M>N$}   & {\color[HTML]{4C4C4C} $B=M^{iii}$}                   & \cellcolor[HTML]{E5F5E0}{\color[HTML]{4C4C4C} $B>C$}   & \cellcolor[HTML]{E5F5E0}{\color[HTML]{4C4C4C} $B>N$} \\
{\color[HTML]{4C4C4C} 3}           & {\color[HTML]{4C4C4C} Quantity}            & {\color[HTML]{4C4C4C} H3}            & {\color[HTML]{4C4C4C} 3}            & \cellcolor[HTML]{E5F5E0}{\color[HTML]{4C4C4C} $M > C$} & \cellcolor[HTML]{E5F5E0}{\color[HTML]{4C4C4C} $C>N$}       & \cellcolor[HTML]{E5F5E0}{\color[HTML]{4C4C4C} $M>N$}   & {\color[HTML]{4C4C4C} $B=M^{iii}$}                   & \cellcolor[HTML]{E5F5E0}{\color[HTML]{4C4C4C} $B>C$}   & \cellcolor[HTML]{E5F5E0}{\color[HTML]{4C4C4C} $B>N$} \\
{\color[HTML]{4C4C4C} 4}           & {\color[HTML]{4C4C4C} Quantity}            & {\color[HTML]{4C4C4C} H6}            & {\color[HTML]{4C4C4C} 4}            & \cellcolor[HTML]{E5F5E0}{\color[HTML]{4C4C4C} $M =C$}  & \cellcolor[HTML]{E5F5E0}{\color[HTML]{4C4C4C} $C=N$}       & \cellcolor[HTML]{E5F5E0}{\color[HTML]{4C4C4C} $M=N$}   & \cellcolor[HTML]{E5F5E0}{\color[HTML]{4C4C4C} $B=M$} & \cellcolor[HTML]{E5F5E0}{\color[HTML]{4C4C4C} $B=C$}   & \cellcolor[HTML]{E5F5E0}{\color[HTML]{4C4C4C} $B=N$} \\
{\color[HTML]{4C4C4C} 5}           & {\color[HTML]{4C4C4C} Context}             & {\color[HTML]{4C4C4C} H2}            & {\color[HTML]{4C4C4C} 1}            & {\color[HTML]{4C4C4C} $M<C$}                       & \cellcolor[HTML]{E5F5E0}{\color[HTML]{4C4C4C} $C=N$}       & \cellcolor[HTML]{E5F5E0}{\color[HTML]{4C4C4C} $M=N$}   & \cellcolor[HTML]{E5F5E0}{\color[HTML]{4C4C4C} $B=M$} & \cellcolor[HTML]{E5F5E0}{\color[HTML]{4C4C4C} $B=C$}   & \cellcolor[HTML]{E5F5E0}{\color[HTML]{4C4C4C} $B=N$} \\
{\color[HTML]{4C4C4C} 6}           & {\color[HTML]{4C4C4C} Context}             & {\color[HTML]{4C4C4C} H4}            & {\color[HTML]{4C4C4C} 2}            & \cellcolor[HTML]{E5F5E0}{\color[HTML]{4C4C4C} $M < C$} & \cellcolor[HTML]{E5F5E0}{\color[HTML]{4C4C4C} $C>N^{ii}$} & {\color[HTML]{4C4C4C} $M=N^{ii}$}                      & \cellcolor[HTML]{E5F5E0}{\color[HTML]{4C4C4C} $B>M$} & \cellcolor[HTML]{E5F5E0}{\color[HTML]{4C4C4C} $B<C$}   & {\color[HTML]{4C4C4C} $B>N^{ii}$}                    \\
{\color[HTML]{4C4C4C} 7}           & {\color[HTML]{4C4C4C} Context}             & {\color[HTML]{4C4C4C} H4}            & {\color[HTML]{4C4C4C} 3}            & \cellcolor[HTML]{E5F5E0}{\color[HTML]{4C4C4C} $M < C$} & \cellcolor[HTML]{E5F5E0}{\color[HTML]{4C4C4C} $C>N^{ii}$}  & {\color[HTML]{4C4C4C} $M=N^{ii}$}                      & {\color[HTML]{4C4C4C} $B=M^{i,iii}$}                 & \cellcolor[HTML]{E5F5E0}{\color[HTML]{4C4C4C} $B < C$} & {\color[HTML]{4C4C4C} $B=N^{i,iii}$}                 \\
{\color[HTML]{4C4C4C} 8}           & {\color[HTML]{4C4C4C} Context}             & {\color[HTML]{4C4C4C} H5}            & {\color[HTML]{4C4C4C} 4}            & \cellcolor[HTML]{E5F5E0}{\color[HTML]{4C4C4C} $M < C$} & {\color[HTML]{4C4C4C} $C=N^{i}$}                           & \cellcolor[HTML]{E5F5E0}{\color[HTML]{4C4C4C} $M < N$} & {\color[HTML]{4C4C4C} $B= M^{i}$}                    & \cellcolor[HTML]{E5F5E0}{\color[HTML]{4C4C4C} $B<C$}   & \cellcolor[HTML]{E5F5E0}{\color[HTML]{4C4C4C} $B=N$} \\
{\color[HTML]{4C4C4C} 9}           & {\color[HTML]{4C4C4C} Accuracy}            & {\color[HTML]{4C4C4C} H7}            & {\color[HTML]{4C4C4C} all}          & {\color[HTML]{4C4C4C} $M = C$}                    & {\color[HTML]{4C4C4C} $C=N^{i}$}                           & \cellcolor[HTML]{E5F5E0}{\color[HTML]{4C4C4C} $M < N$} & {\color[HTML]{4C4C4C} $B= M^{i}$}                    & {\color[HTML]{4C4C4C} $B=C$}                      & {\color[HTML]{4C4C4C} $B<N^{i}$}                     \\ 
\hline          
\end{tabular}
\end{table*}

Table \ref{tab:kruskal} depicts the results of the Kruskal-Wallis test applied to the distributions of the dependent variables for each day/ over all days of the four treatment groups. Moreover, Table \ref{tab:post_hoc_days} and \ref{tab:post_hoc_all} illustrate the post-hoc analysis that applies the Conover-Iman test for those days which exhibit significant differences in the Kruskal-Wallis analysis. Moreover, Figure \ref{fig:cdf} depicts the cumulative distribution for each treatment group and Figure \ref{fig:interaction_plots} shows the interactions among the treatments for the analyzed dependent variables.
 
 In the following, the observations for each dependent variable are illustrated in detail.

\subsubsection{Quantity}
The treatment group behaviors for the quantity variable are significantly different for Day~2 and Day~3 (Table \ref{tab:kruskal}). Considering the post-hoc analysis (Table \ref{tab:post_hoc_days}), it is possible to determine that for both days, all treatment group pairs are significantly different, except for the B-M (both token incentives-money token incentive) pair. The CDF plot illustrates this observation (Figure \ref{fig:cdf}): The M and B groups have a similar higher probability to provide more answers when compared to the C (context token incentive) and N (control group) groups (in this order). Moreover, M and B distributions show two peaks, one around 60 answers and one around 150 answers, the latter being the maximum number of answers for which a payment is received on a given day (Section \ref{sec:exp_treatment}). These peaks are more clear visible on Day 3 and are stronger for the B group when compared to the M group. Moreover, the CDF plot for Day 3 (Figure \ref{fig:quant_day3_cdf}) indicates a tendency for money token receivers to answer a higher number of questions.

The plots in Figure \ref{fig:interaction_plots} illustrate the median interaction effects.
Similarily to the Kruskal-Wallis test, on Day 1 and Day 4 no effect of the incentives are identified (Figure \ref{fig:quantity_day1_interaction} and \ref{fig:quantity_day4_interaction}). On days~2 and 3 (Figure \ref{fig:quantity_day2_interaction} and \ref{fig:quantity_day3_interaction}), both incentives result in an increase of questions answered when compared to the control group, whereby the money token leads to a considerably stronger increase than the context token. Moreover, at Day~3 an interaction is observed: When compared to the money group, the context token dampens the effect of the money token in the B group resulting in fewer questions answered. 

\subsubsection{Contextualization}
In contrast to quantity, the treatment group behaviors are significantly different for all four days (Table \ref{tab:post_hoc_days}). The interaction plot on Day 1 (Figure \ref{fig:context_day1_interaction}) illustrates how the context token treatment resulted in a higher number of contextualizations. The CDF plot on Day 1 (Figure \ref{fig:context_day1_cdf}) depicts a similar distribution of the treatment groups with a higher tendency of the context group to provide more contextualizations. 
On Day~2, all groups except the M-N pair are significantly different (Table \ref{tab:post_hoc_days}).  Nevertheless, on Day 3 no differences between the B-N and B-M pairs are observed any longer. An opposing trend is observed in the M-N and B-C pairs where the p-values become smaller over the two days (Table \ref{tab:post_hoc_days}). 

The CDF plots for Day 2 and 3 (Figure \ref{fig:context_day2_cdf} and \ref{fig:context_day3_cdf}) illustrate these trends: On Day 2, the B group distribution is close to the C distribution. Nevertheless, on Day 3 it more closely resembles the M group distribution, where most individuals provide few contextualizations and few individuals many. Moreover, the difference between the M and N groups becomes stronger for individuals that provide few contextualizations.

The interaction plots for Day 2 and day 3 (Figure \ref{fig:context_day2_interaction} and \ref{fig:context_day3_interaction}) illustrate an interaction effect between the money and context token resulting in fewer contextualizations when compared to the context token alone. In contrast to the money token, no trend in the interaction is observed over the four days.

After removing the incentives on Day 4, the pairs B-C and C-M and M-N show a significantly different behavior (Table \ref{tab:post_hoc_days}). This is in contrast to the observation for the quantity variable, where no distinct behavior on Day 4 is identified. 
No significant difference between treatments with the context token and the control group is identified. The CDF plot for Day 4 (Figure \ref{fig:context_day4_cdf}) illustrates the similarity between the B-M groups, and respectively between the C-N groups and the difference between each of these pairs. 
The interaction plot (Figure \ref{fig:context_day4_interaction}) illustrates how the C group provides the most contextualizations on Day 4, followed by the control group and then the other two groups. 

\subsubsection{Accuracy}

A significant difference among the group behaviors for the accuracy is identified over all four days (Table~\ref{tab:kruskal}). Table~\ref{tab:post_hoc_all} indicates that this difference originates from the pairs B-N and M-N. Nevertheless, the p-values of the pairs B-C and M-C are also almost significant (p-value = 0.052). These differences are illustrated in the CDF-plot (Figure \ref{fig:accurarcy_cdf}) which depicts higher probabilities for the C and N group to reach higher accuracy values when compared to the other two groups. Figure \ref{fig:accuracy_interaction} shows that the control group reaches the highest accuracy in their answers, followed by the context, both and money groups. 



\section{Discussion}
\label{sec:discussion}

\subsection{Results}

Table \ref{tab:findings} illustrates the findings from the Kruskal-Wallis and post-hoc analysis with regard to the hypotheses (Section~\ref{sec:hypotheses_formulation}). The results inform an adjustment to the assumptions (Section~\ref{sec:assumptions}) that were utilized in the formulation of these hypotheses: 

I) It was assumed that the context token has a positive impact on both the intrinsic and extrinsic motivation (Assumption \ref{ass:context_reputation} and \ref{ass:context_extrinsic} in Section \ref{sec:hypotheses}). Yet, the findings provide evidence that this token has overall only a small positive or negligible impact on intrinsic motivation. This would explain the parity of the C group and N group for the context characteristic on Day 4 (ID 8/Column b in Table \ref{tab:findings}): No incentives are applied on that day, thus the extrinsic motivation is equally zero for both groups and only the intrinsic motivation defines the contextualization behavior. However, the median number of contextualizations on Day 4 (Figure \ref{fig:context_day4_interaction}) is higher for the C group when compared to the N group indicating a small positive impact that is also illustrated by the CDF plot (Figure \ref{fig:context_day4_cdf}).
The neglible impact is also illustrated in the parity between the B and M group on Day 4 (ID 8/ Column d in Table \ref{tab:findings}). The money token reduced the intrinsic motivation in both groups and because the context token did not offset this negative impact, it is the same for both groups on Day 4. 
Moreover, the parity of these pairs for the accuracy characteristic (ID 9/Column b and d in Table \ref{tab:findings}) can be explained thus: Since accuracy is mainly impacted by intrinsic motivation, which following the previous considerations is equally pronounced between the M and B groups, the group behaviors are equal. It also explains the inequality between the B and N group (ID 9/ Column f in Table \ref{tab:findings}): Since the intrinsic motivation is reduced in the B group due to the money token, and the context token does not have a significant positive impact on intrinsic motivation, the intrinsic motivation in the B group is lower than in the N group and thus the shared information is of lesser accuracy.  

II) Extrinsic motivation has a considerable impact on the context characteristic. i) This explains the parity between the M and N group for Days 2 and 3 (ID 6,7/Column c in Table \ref{tab:findings}). Although the intrinsic motivation is reduced due to the money token, it is replaced by the extrinsic motivation stemming from this incentive resulting in a similar contextualization-sharing behavior. Consequently at Day 4, when the incentive is removed, the M group shares fewer contextualizations (ID 8/Column c in Table \ref{tab:findings}). ii) It also explains the inequality between the B and N group on Day 2 (ID 6/Column f in Table \ref{tab:findings}). In contrast to the comparison between the M and N group, the context token in the B group adds to the extrinsic motivation such that the decrease in intrinsic motivation is exceeded resulting in a greater motivation to share contextualizations when compared to the N group. iii) Moreover, following the same arguments, it also explains the inequality between the B and M group on Day 2 (ID 6/Column d in Table \ref{tab:findings}). iv) Finally, it also explains the inequality between the C and N group on Days 2 and 3 (ID 6,7/Column b in Table \ref{tab:findings}).


III) In contrast to Assumption \ref{ass:no_interaction}, interactions between the money and context token incentive are observed. i) For Day 2, the B group shares more contextualizations than the monetary group (ID 6 Column d in Table \ref{tab:findings}). Nevertheless, on Day 3 no difference is observed (ID 7 Column d in Table \ref{tab:findings}), which indicates that over time the two tokens interact with each other, thereby decreasing their impact on the users' motivation. ii) This might also explain the parity between these groups for the quantity of information shared on Day 2 and 3 (ID 2,3/ Column d in Table \ref{tab:findings}): Both tokens interfere such that their combined impact on users' motivation does not differ from a single token incentive. Furthermore, the interaction plot (Figure \ref{fig:interaction_plots}) even indicates a lower positive impact of the combined incentives when compared to the single money token incentive. In addition, the plot indicates that this interaction becomes stronger over the four days, which is also illustrated by the CDF plot.
iii) This interaction also explains the shift from inequality to equality for the B and N group on Days 2 and 3 (ID 6,7 Column f in Table \ref{tab:findings}).

The findings further provide evidence that the money token crowds out intrinsic motivation.
The interaction plots and CDF plots for the accuracy indicate a crowding-out of intrinsic motivation by the context token. Nevertheless, according to the Kruskal-Wallis test and its post-hoc analysis, these latter differences are not significant. In particular, for the number of contextualizations the context token even has a positive impact after incentivization ends, which might be explained by an internalization of the incentive (Section \ref{sec:std_theory}) for this information dimension.
Finally, a time effect is present in both, single- and multiple-token scenarios, which indicates that the behavioral change can vary over time.

\subsection{Implications}

\subsubsection{One-dimensional token systems}
\label{sec:one_dimensional_token_system}
The internalization effect of the context token on contextualization actions after incentivization ends illustrates a potential advantage of blockchain-based cryptoeconomic incentives when compared to traditional approaches utilizing monetary incentives. The intrinsic motivation of users might be impacted positively by internalizing these incentives (Section \ref{sec:std_theory}), thus resulting in an improvement of performance measured in the amount of contextualizations provided, even after the incentivization period ends.
However, the findings also indicate that this utility token induces a worse performance in the accuracy of shared data when compared to the control group (Figure \ref{fig:accurarcy_cdf} and \ref{fig:accuracy_interaction}). Therefore, the identified internalization might be limited to information dimensions that are directly incentivized by a utility token. 

Moreover, the findings indicate that stable coins such as the utilized money token crowd out intrinsic motivation, which resulted in this work in a reduction of information quality measured in accuracy and contextualization.

In order to design effective incentives, future work should evaluate the token designs and scenarios under which internalization or crowding-out are observed. In particular, as internalization or crowding-out can vary between different performance measures (e.g. contextualization and accuracy), one is advised to carefully evaluate all impacts a token might have before using it in real-world applications.

\subsubsection{Multi-dimensional token systems}
The findings provide evidence that applying multiple token-based incentives simultaneously can result in a combined improvement of several information characteristics (e.g., as shown for quantity and contextualization) and could therefore improve system performance when compared to a scenario where a single token is utilized.
Nevertheless, the identified interaction effect between the two tokens of this paper indicates that designing multi-token systems is a non-trivial task that has implications for systems in which the application of multiple tokens is considered.  
In particular, positive and negative impacts of tokens on human behavior may not simply add up. The findings also show that these effects may only become apparent over time. Thus, it is necessary to carefully analyze the interdependencies between combinations of tokens in longitudinal studies before they are utilized in real-world systems. The results of simulations and formal analyses of multi-dimensional token systems are limited if they do not consider these token interactions.  


\subsubsection{(Ethical) risks}
Considering the observation, that the current big data paradigm is not challenged by a lack of data~\cite{helbing2015digital}, but contextualized and accurate information, the findings of this paper raise the question if incentives in the form of blockchain-based tokens should be applied at all to motivate individuals of communities to share information. Such incentives may result in a further increase in quantity of collected information while reducing its quality (e.g. accuracy). 
 

However, incentives might work differently in data-sharing scenarios where the quality of shared data under different incentivizations is determined by decisions users take a priori, which are then posteriori executed by an artificial intelligence, as studied by \citet{pournaras2016self,asikis2020optimization} with a computational methodology for privacy-utility decisions. 
Yet, neither the user acceptance of these decisions nor the impact they have on the trust of users for decision-support systems have been studied. Therefore, unknown effects could be present in these scenarios that bias users' behavior and which limit the generalizability of findings from such simulations to real-world situations. Furthermore, because users have to perform decisions a priori in these scenarios, such approaches might fail to capture the unique situational domain knowledge users possess or their creativity and intuition which is required for some application domains such as the customer feedback provision analyzed in this paper.


Increasingly, token incentives are applied in various application domains of society such as construction~\cite{hunhevicz2020incentivizing}, health~\cite{jung2021mechanism}, Covid-19 prevention measures~\cite{manoj2020incentive}, electricity production and consumption~\cite{wittekcrypto}, car sharing~\cite{kim2018blockchain,valavstin2019blockchain}, alleviating traffic congestion~\cite{aung2020t}, book-keeping~\cite{cai2019analysis}, decentralized access-control systems~\cite{gan2020token}, or waste reduction~\cite{pardi2021chemical}. Nevertheless, behavioral traits stemming from intrinsic motivation, such as creativity, joy, self-determination, purpose, and endurance may be important for some of those application domains which could be crowded out by these cryptoeconomic incentives and thus would result in reduced performance. For instance, endurance~\cite{hackmann2014social} and creativity~\cite{thornton2014climate} have been identified as important factors for addressing climate change.

Moreover, this increasing tokenization of areas of life that have not been tokenized before could reduce social relations and human interactions to transactions within a market-driven economy~\cite{pazaitis2017blockchain}. This might be in opposition to values that stakeholders in these systems hold~\cite{pazaitis2017blockchain}. 
In addition, it has been found that the measurement act itself, which is required in tokenization for the quantifying and proving of actions~\cite{dapp2021finance}, can reduce intrinsic motivation and thus creativity and endurance in individuals~\cite{etkin2016hidden}.

In addition, the identified effects of this work question the assumptions of controversial token systems in the form of social credit systems~\cite{creemers2018china,liang2018constructing}, which are also discussed in Western democracies as tools for managing society~\cite{bmbf_2020}: Centrally designing and introducing token incentives may fail due to unkown, crowding-out or interdependent effects that may cascade over time. Considering the large and complex design space of DLT systems~\cite{ballandies2021decrypting}, an iterative, local, and community-driven approach utilizing the wisdom of crowds and self-organization for token designs as illustrated in \citet{dapp2021finance} might be the way to proceed in designing stable token systems. In particular, these principles have been found to enable communities to mitigate the tragedy of the commons and successfully share and maintain a common resource~\cite{baur2021measures}.

Thus, before applying token incentives in an application scenario, this author suggests rigorously considering the values of all stakeholders in the system construction process and analyzing whether applying such incentives could crowd out intrinsic motivation in the scenario under consideration. Only then should the system be iteratively constructed in scenarios that are locally bound. For this, the methodology of this paper in combination with value-sensitive design~\cite{friedman2013value,van2015handbook} and iterative design science research~\cite{hevner2004design,vom2020special} methodologies can be applied, as demonstrated for token-based blockchain systems by \citet{ballandies2021finance,ballandies2022improving}.

\subsubsection{Limitations}

The experiment facilitates realism while enabling the laboratory-like testing of hypotheses~\cite{pournaras2022how}. Due to the realism sought, not all influences on users' information-sharing behavior could be controlled for, which may reduce the quality of the measurements and findings. 
In particular, the questions asked are formulated by the library organization which had a real business interest in the answers. Thus, the questions are not standardized and hence, some of the questions might be more difficult to answer. This could introduce bias to quality characteristics such as accuracy and may have resulted in the lower differentiation between treatment groups in this characteristic (ID~9 in Table~\ref{tab:findings}). Furthermore, the accuracy was summarized over the four days. As a result, there is a lack of a granular daily view on the impact of token incentives on this characteristic. 

\subsubsection{Impact}
\label{sec:impact}
The realistic setup of the experiment illustrates and underlines the importance of the findings of this paper for real-world organizations and communities. The identification of significant positive and negative effects of both token incentives on human sharing behavior and their observed interactions provide evidence that such effects are present in real-world sharing scenarios and should therefore be analyzed and evaluated by organizations and communities before they are applied in their use cases. In particular, a token design may not be robust, with use of the token having a different impact than intended~\cite{ballandies2021financebook}. The methodology of this paper can be applied to analyze such effects in real-world systems. 

The identified effects (interaction, internalization, time, and crowding-out) inform the Token Engineering and Token Economics community in the design of stable cryptoeconomies. Currently, methodologies in these fields mainly rely on game theory, mechanism design, and simulations~\cite{barreiro2019blockchain,kim2021token,zhang2019engineering,khamisa2021token,laskowski2020evidence,tan2020economics}. Nevertheless, none of these approaches considers the identified effects of this work on human behavior in their assumptions. Consequently, including these effects could improve the correspondence of findings from these methodologies with reality. Thus, this paper demonstrates the importance of behavioral experiments in the field of Token Engineering and Token Economics. 

In addition, this work illustrates the usability of self-determination theory to test hypotheses of token designs on human behavior.

\section{Conclusion and Outlook}
\label{sec:conclusion}
This work evaluates the combined impact of multiple cryptoeconomic incentives in the form of blockchain-based tokens on human information-sharing behavior. By utilizing a rigorous experimental methodology with a 2x2 factorial design involving 132 participants, the impact is evaluated in a real-world information-sharing scenario involving a major Swiss organization and its customers. The identified interaction effect between the tokens and the potential crowding-out of intrinsic motivation by these cryptoeconomic incentives are important for researchers and practitioners to consider because they indicate that designing multi-token systems is a non-trivial task: The impact of individual token incentives on human behavior are not independent from each other and a token design might not be sufficiently robust, with the impact of the token possibly differing from the intended effect. These impacts have to be considered when implementing, simulating or mathematically analysing token economies as presented in the Discussion (Section \ref{sec:impact}). In particular, they inform the assumptions taken in theoretical models, validate their accuracy, and may thus facilitate their improved connection with reality. 
Therefore, the methodology of this paper and the identified effects might be of use for organizations and communities that intend to apply (multiple) token incentives.

The results point to various avenues for future research. i) Since information quality is a multi-dimensional concept (Section \ref{sec:variables_measures}) and the impact of token incentives can vary between those dimensions (Section \ref{sec:one_dimensional_token_system}), the impact of the chosen token incentives on other operationalizations of quality than accuracy or contextualization can be evaluated to further quantify the impact of these tokens on human information-sharing behavior. ii) In general, considering the broad design space of tokens and blockchain systems, the impact of further instances of cryptoeconomic token incentives should be evaluated in experiments to identify conditions and scenarios that are impaired or benefit from the introduction of cryptoeconomic incentives. iii) Due to the identified interaction effect and the complexity of potential system layouts, evaluating all these combinations in experimental setups might not be feasible. Thus, simulations should be employed to identify areas of interest in the design space, which, in a second step, are investigated in experiments. Modeling the determined effects of this work as emergent phenomena of a complex system could be a promising approach for these simulations. 
Finally, machine learning methods such as k-means or hierarchical clustering could be utilized to identify hidden patterns in the data that may impact human sharing behavior under incentivization.

Thus, to conclude, further research by the cryptoeconomics community is required to identify why, how, and in which situations cryptoeconomic incentives should be applied.

\begin{landscape}
\begin{table}[]\caption{p-values obtained from the Kruskal-Wallis test when comparing the different treatment groups for the five dependent variables. Levels identifying significant differences among the treatment groups distributions: $\leq$ 0.001 ***, $\leq$ 0.01 **, $\leq$ 0.5 *.}\label{tab:kruskal}
\begin{tabular}{cccc}\hline \\
\textbf{Day} & \multicolumn{3}{c}{\textbf{p-value}}                                          \\
             & \textit{Quantity} & \textit{Context.} & \textit{Accuracy}  \\\hline \\
\textit{1}   & 0.957            & 0.029*           & -                                    \\
\textit{2}   & 0***             & 0***             & -                                    \\
\textit{3}   & 0***             & 0***             & -                                    \\
\textit{4}   & 0.795            & 0***             & -                                    \\
\textit{All} & 0***             & 0***             & 0.021*                       \\ \hline  
\end{tabular}
\end{table}

\begin{table}[] \caption{Daily p-values of Conover-Iman post-hoc analysis for the significant values of the Kruskal-Wallis test (Table \ref{tab:kruskal}) for the quantity and contextualization variables. 
}\label{tab:post_hoc_days} 
\begin{tabular}{cccccc|cccc|cccc|cccc} \hline 
\multicolumn{1}{l}{}           & \multicolumn{1}{l}{} & \multicolumn{4}{c}{\textbf{Day 1: No incentives}}                & \multicolumn{4}{c}{\textbf{Day 2: Token incentives}}                & \multicolumn{4}{c}{\textbf{Day 3: Token incentives}}                & \multicolumn{4}{c}{\textbf{Day 4: No incentives}}                \\
\multicolumn{1}{l}{}           & \textbf{Treat.}      & \textit{B} & \textit{C} & \textit{M} & \textit{N} & \textit{B} & \textit{C} & \textit{M} & \textit{N} & \textit{B} & \textit{C} & \textit{M} & \textit{N} & \textit{B} & \textit{C} & \textit{M} & \textit{N} \\ \hline \\
\multirow{4}{*}{\textbf{Quantity}} & \textit{B}           & -          & -          & -          & -          & 1          & 0          & 0.683      & 0          & 1          & 0.005      & 0.48       & 0          & -          & -          & -          & -          \\
                               & \textit{C}           & -          & -          & -          & -          & 0          & 1          & 0          & 0.001      & 0.005      & 1          & 0.001      & 0.005      & -          & -          & -          & -          \\
                               & \textit{M}           & -          & -          & -          & -          & 0.683      & 0          & 1          & 0          & 0.48       & 0.001      & 1          & 0          & -          & -          & -          & -          \\
                               & \textit{N}           & -          & -          & -          & -          & 0          & 0.001      & 0          & 1          & 0          & 0.005      & 0          & 1          & -          & -          & -          & -          \\ \hline 
\multirow{4}{*}{\textbf{Context.}} & \textit{B}           & 1          & 0.204      & 0.712      & 0.644      & 1          & 0.049      & 0.02       & 0.02       & 1          & 0.001      & 0.098      & 0.393      & 1          & 0.007      & 0.289      & 0.289      \\
                               & \textit{C}           & 0.204      & 1          & 0.035      & 0.712      & 0.049      & 1          & 0          & 0          & 0.001      & 1          & 0          & 0          & 0.007      & 1          & 0          & 0.227      \\
                               & \textit{M}           & 0.712      & 0.035      & 1          & 0.181      & 0.02       & 0          & 1          & 0.926      & 0.098      & 0          & 1          & 0.395      & 0.289      & 0          & 1          & 0.041      \\
                               & \textit{N}           & 0.644      & 0.712      & 0.181      & 1          & 0.02       & 0          & 0.926      & 1          & 0.393      & 0          & 0.395      & 1          & 0.289      & 0.227      & 0.041      & 1          \\ \hline 

\end{tabular}
\end{table}

\begin{table}[]\caption{p-values of Conover-Iman post-hoc analysis for the significant values of the Kruskal-Wallis test (Table \ref{tab:kruskal}) for the accuracy variable over all days. 
}\label{tab:post_hoc_all} 
\begin{tabular}{cc|cccc|cccc|}\hline 
\multicolumn{1}{l}{}                    &                 & \multicolumn{4}{c}{\textbf{Accurcacy}}                                                                                                       \\
\multicolumn{1}{l}{}                    & \textbf{Treat.} & \textit{B} & \textit{C} & \textit{M} & \textit{N}  \\ \hline \\
\multirow{4}{*}{\textbf{Over all days}} & \textit{B}      & 1          & 0.052      & 0.957      & 0.004                                \\
                                        & \textit{C}      & 0.052      & 1          & 0.052      & 0.603                               \\
                                        & \textit{M}      & 0.957      & 0.052      & 1          & 0.004                               \\
                                        & \textit{N}      & 0.004      & 0.603      & 0.004      & 1            \\ \hline                          
\end{tabular}
\end{table}

\end{landscape}

\appendices

\section*{Ethical approval}


\section*{Acknowledgment}

I thank Evangelos Pournaras, Dirk Helbing, Claudio Tessone and Carina I. Hausladen for their valuable comments. Moreover, I would like to thank Stefan Wehrli and the rest of the ETH DeSciL staff members for their support in conducting the experiment. Furthermore, I thank Maximiliane Okonnek and the ETH library lab team for providing the financial support of the experiment, the access to their infrastructure, and their valuable feedback. Finally, I thank Lewis J. Dale for his assistance in the proofreading of this paper.

\bibliographystyle{unsrtnat}
\bibliography{access}

\end{document}